\documentclass[proof]{pasj00}
\draft

\begin{document}
\SetRunningHead{Author(s) in page-head}{Running Head}
\Received{2003 May 7}
\Accepted{2003 July 8}

\title{Design and Performance of the Wide-Field X-Ray Monitor on Board the 
    High-Energy Transient Explorer 2}


\author{

Yuji     \textsc{Shirasaki}, \altaffilmark{1,2}
Nobuyuki \textsc{Kawai},     \altaffilmark{2,3}
Atsumasa \textsc{Yoshida},   \altaffilmark{2,4}
Masaru   \textsc{Matsuoka},  \altaffilmark{5}\\
Toru     \textsc{Tamagawa},  \altaffilmark{2}
Ken'ichi \textsc{Torii},     \altaffilmark{2}
Takanori \textsc{Sakamoto},  \altaffilmark{2,3}
Motoko   \textsc{Suzuki},    \altaffilmark{3}\\
Yuji     \textsc{Urata},     \altaffilmark{2,3}
Rie      \textsc{Sato},      \altaffilmark{3}
Yujin    \textsc{Nakagawa},  \altaffilmark{4}
Daiki    \textsc{Takahashi}, \altaffilmark{4}\\

Edward E. \textsc{Fenimore}, \altaffilmark{6}
Mark      \textsc{Galassi},  \altaffilmark{6}

Donald Q. \textsc{Lamb},     \altaffilmark{7}
Carlo     \textsc{Graziani}, \altaffilmark{7}\\
Timothy Q. \textsc{Donaghy},  \altaffilmark{7}

Roland    \textsc{Vanderspek},\altaffilmark{8}\\

Makoto    \textsc{Yamauchi},  \altaffilmark{9}
Kunio     \textsc{Takagishi}, \altaffilmark{9}
and
Isamu     \textsc{Hatsukade}  \altaffilmark{9}
}

\altaffiltext{1}{National Astronomical Observatory of Japan, Osawa, Mitaka, Tokyo 181-8588}
\email{yuji.shirasaki@nao.ac.jp}
\altaffiltext{2}{RIKEN, Hirosawa, Wako, Saitama 351-0198}
\altaffiltext{3}{Department of Physics, Tokyo Institute of Technology, Meguro-ku, Tokyo 152-8551}
\altaffiltext{4}{Department of Physics, Aoyama Gakuin University, Sagamihara, Kanagawa 229-8558}
\altaffiltext{5}{NASDA, Tsukuba, Ibaraki 304-8505}
\altaffiltext{6}{Los Alamos National Laboratory, Los Alamos, NM 87545, USA}
\altaffiltext{7}{Department of Astronomy and Astrophysics,
                 University of Chicago, Chicago, IL 60637, USA}
\altaffiltext{8}{MIT/CSR  77 Massachusetts Avenue, Cambridge, MA 02139, USA }
\altaffiltext{9}{Faculty of Engineering, Miyazaki University, 
                 Miyazaki 889-2192}





\maketitle

\begin{abstract}
     The Wide-field X-ray Monitor (WXM) is one of the scientific
     instruments carried on the High Energy Transient Explorer 2 (HETE-2)
     satellite launched on 2000 October 9.
     HETE-2 is an international mission consisting of a
     small satellite dedicated to provide broad-band observations and
     accurate localizations of gamma-ray bursts (GRBs). 
     A unique feature of this mission is its capability to determine and
     transmit GRB coordinates in almost real-time through the burst
     alert network.
     The WXM consists of three elements: four identical Xe-filled
     one-dimensional position-sensitive proportional counters,
     two sets of one-dimensional coded apertures, and 
     the main electronics.
     The WXM counters are sensitive to X-rays between 2 keV and 25 keV
     within a field-of-view of about 1.5 sr, with a total detector area
     of about 350 cm$^2$. 
     The in-flight triggering and localization capability can
     produce a real-time GRB location of several to 30 arcmin
     accuracy, with a limiting sensitivity of $10^{-7}$ erg cm$^{-2}$. 
     In this report, the details of the mechanical structure, 
     electronics, on-board software, ground and in-flight
     calibration, and in-flight performance of the WXM are discussed.
\end{abstract}

\section{Introduction}
\label{sect:intro}

   The origin and nature of Gamma-ray bursts (GRBs) has been one of the
outstanding mysteries of astrophysics since the phenomenon was discovered
in the 1970s.
   The discovery of GRB afterglows by the BeppoSAX X-ray
satellite~\citep{1997Natur.387..783C} produced a breakthrough in the
study of GRB.
   Since the discovery, follow-up observations of many GRB afterglows have
been successfully performed.  The observations have furnished strong
evidence that at least the long class of GRB originates from explosions of
super-massive stars at cosmological distances.
   The HETE-2~\citep{wh.2003.grr} is the first satellite dedicated to
GRB observations and is intended to provide the astronomical
community with prompt, high-accuracy GRB positions --- a few tens of
arcminutes to a few tens of arcseconds, with delay times of 10 s to
a few hours.
   This capability not only increases the number of 
detected GRB counterparts, but also provides an opportunity for multi-wavelength
observations of GRBs in their very early phase.  Such observations were
nearly impossible prior to the HETE-2 era.

   The HETE-2 instrument complement consists of three scientific instruments:
the French Gamma Telescope (FREGATE), the Wide-field X-ray Monitor (WXM),
and the Soft X-ray Camera (SXC).
   FREGATE~\citep{wh.2003.atteia} consists of four identical
scintillation detectors.  It is sensitive to photons with energies from 6 to
400 keV.
   It is responsible for triggering and spectroscopy in the hard X-ray
and Gamma-ray energy ranges.
   Its field of view is 70$^{\circ}$ (half width at zero maximum)
and its total detection area is 160 cm$^2$.
   The SXC~\citep{wh.2003.joel} is a 1-D coded mask system using MIT-LL
CCID-20's as the detecting elements, and is sensitive to photons with
energies from 500 eV to 14 keV, with 2\% resolution at 6 keV and with a 
position resolution of 10$''$ for bright bursts (1 Crab; 10 s).
   SXC's total detection area is 75 cm$^2$ and its field of view is 
$65^{\circ}$ (full width at zero maximum).
   The WXM, on the other hand, has a detection area of 352 cm$^2$ and a field of
view of $80^{\circ}$ (FWZM).
   Thus, the WXM is more sensitive to weak bursts than the SXC, and
the two X-ray instruments play a complementary role
to each other, balancing detection sensitivity against location resolution.

\section{The WXM Instrument}

\subsection{General Properties of the WXM}
\begin{figure}
   \begin{center}
      \FigureFile(0.5\textwidth,){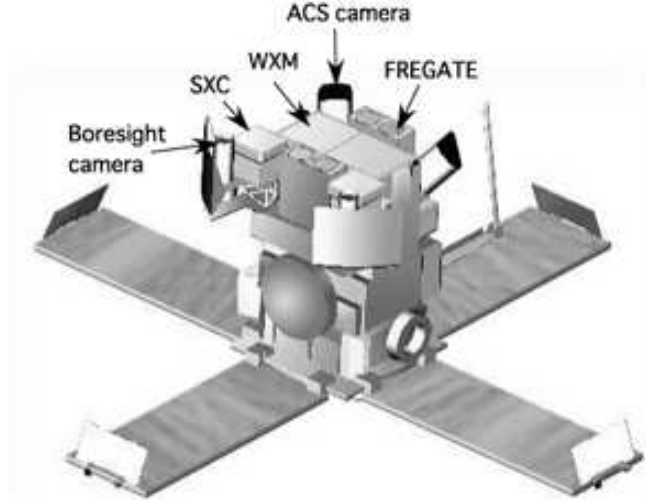}
   \end{center}
   \caption{Schematic drawing of the HETE-2 spacecraft.}
   \label{fig:hete_face}
\end{figure}
   The WXM instrument is located at the center of the spacecraft, as
shown in figure~\ref{fig:hete_face}.
   The HETE-2 spacecraft is in an equatorial orbit, with an inclination
angle
of 2$^\circ$ and a 600 km altitude.  The instruments usually point toward the anti-sun 
direction.  Consequently, the WXM field of view ($80^{\circ} \times
80^{\circ}$) moves along the ecliptic and is obstructed by the earth on
the day side of the orbit.

The WXM consists of two units, the X-camera and the Y-camera. 
Each camera consists of two identical one-dimensional position-sensitive proportional
counters, placed 187~mm beneath a one-dimensional coded-aperture mask.
The PSPCs in the X-camera are referred to as XA and XB, while those in
the Y-camera are referred to as YA and YB.
The orientation of the PSPCs and the mask of the X-camera is perpendicular
to the orientation of the PSPCs and the mask of the Y-camera, so that the X
Y locations of GRBs are determined separately.
Each proportional counter has three anode channels, which are
referred to as XA0, XA1, and XA2 for the case of the XA counter, and similarly
for the other counters.

\begin{figure}
   \centerline{\FigureFile(0.7\textwidth,){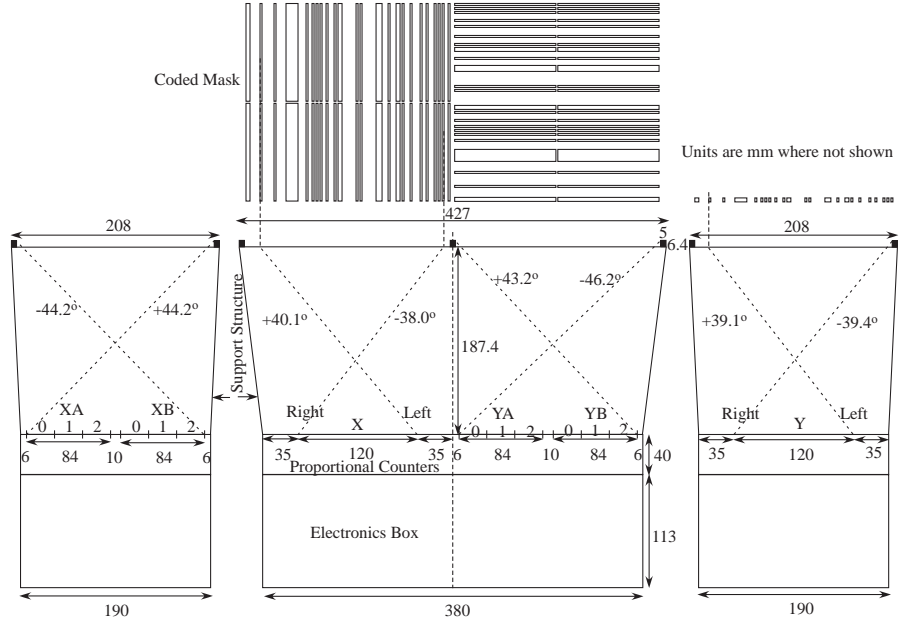}}
   \caption{WXM instrument schematic drawing. 
            A coded mask is placed at 187.37 mm above the proportional 
            counters and supported by an aluminum structure. 
            The field of view is restricted by masks and fasteners, which
            fix the coded mask on the support structure.}
   \label{fig:wxm_drawing}
\end{figure}

   One unit consists of a one-dimensional coded mask and
two 1-D position-sensitive proportional counters (PSPCs) placed
187~mm below the mask.
   The area of each coded mask is twice that of the total
detector area, to ensure a wide field of view.  The masks are supported by an
aluminum support structure.
   Each coded mask consists of a plate of aluminum (0.5~mm thickness) 
plated with gold (50.8 $\mu$m) with a series of slits whose widths are randomly 
varying integer multiples of 2~mm.
   The open fraction of the mask is 0.33.
   A 7.6 $\mu$m thickness aluminum coated
kapton foil is placed in front of the mask, as a thermal shield.
   The WXM instrument schematic drawing is shown in
figure~\ref{fig:wxm_drawing}.
   The field of view of the X unit is geometrically limited
to
$\theta_{x} = -38^{\circ}$ to $+40^{\circ}$ and 
$\theta_{y} = -44^{\circ}$ to $+44^{\circ}$, while that of 
the Y unit is limited to
$\theta_{x} = -46^{\circ}$ to $+43^{\circ}$ and 
$\theta_{y} = -39^{\circ}$ to $+39^{\circ}$,
where $\theta_{x}$ and $\theta_{y}$ are the projection angles measured from the 
vertical direction onto the XZ and YZ plane of the detector coordinate
system, respectively.
   The location of a GRB is determined by independently measuring a set of two
shift distances of the mask pattern shadow on the X and Y detectors.

\subsection{Proportional Counter of the WXM}
   Each PSPC has three anode wires, each composed of a carbon fiber 10~$\mu$m
in diameter and 120 mm in length.  Each PSPC also has four veto wires of gold-plated
tungsten, 20~$\mu$m in diameter.
   The measured resistance of the carbon fibers ranges from 12.0 to 
15.8 k$\Omega$.
   The advantage of using carbon fiber is its ability to withstand 
an electric discharge without damage.
   We have tested this durability using an X-ray generator, and confirmed
that the performance is not changed even after strong X-ray irradiation
comparable in strength to an X4-class solar flare.

\begin{figure}
   \centerline{\FigureFile(0.6\textwidth,){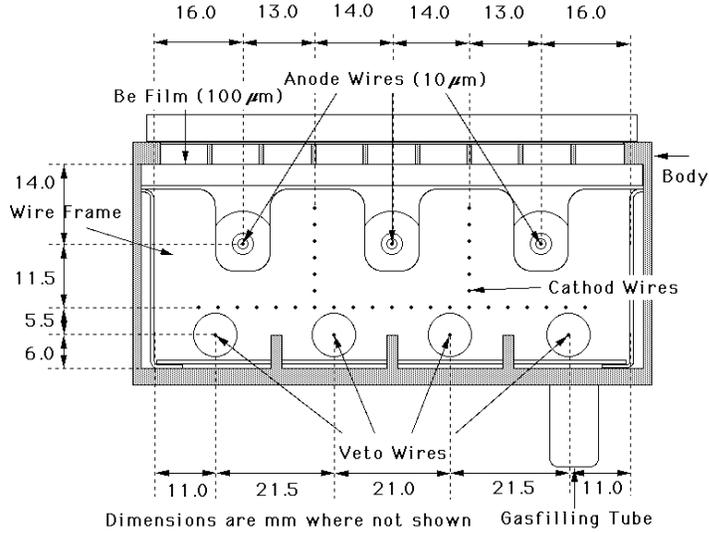}}
         \caption{Cross section of a PSPC of WXM.
         The anode wires in the three upper cells are for 
         X-ray detection and those in the lower cells for rejection
         of charged particle events by the anti-coincidence method. 
         The 120~mm$\times$83.5~mm entrance window
         is sealed by a 100~$\mu$m Be film.}
         \label{fig:pspc_cs}
\end{figure}

\begin{figure}
   \centerline{\FigureFile(0.6\textwidth,){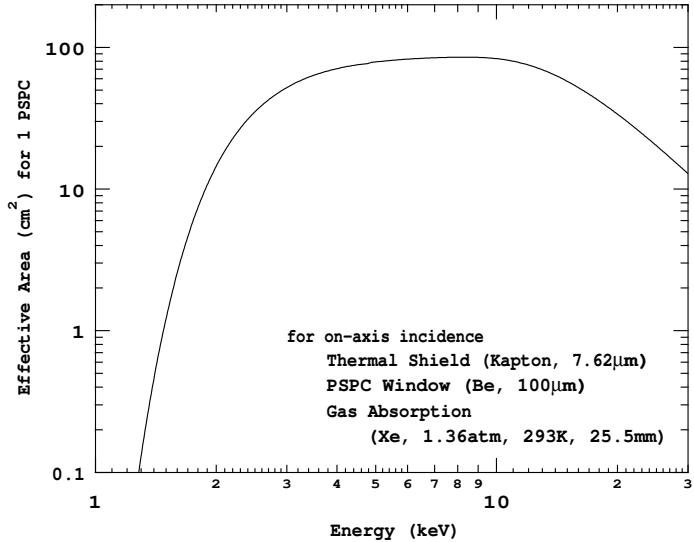}}
   \caption{Calculated effective area of one PSPC for on-axis X-ray
            incidence. Absorption by thermal shield and Be window, and
            stopping probability in the Xe(97\%)+CO$_{2}$(3\%) mixture gas are
	    taken into account. The maximum is 85.4~cm$^{2}$ at 8.3~keV.
            }
   \label{fig:pspc_effective_area}
\end{figure}

   As shown in figure~\ref{fig:pspc_cs}, the inside of the counter is
divided into an upper layer consisting of three anode cells and a lower
veto layer.
   Most of the X-rays are absorbed in the upper cells and produce a
signal on one of the anodes, while charged particles ionize
the Xe gas along their trajectories and produce signals on multiple wires.
   The three upper cells are partitioned by 50~$\mu$m
diameter tungsten cathode wires placed at intervals of 3~mm.
   The counters are filled with xenon (97\%) and carbon dioxide (3\%) at
1.4 atm pressure at room temperature.  They have beryllium entrance
windows of 100~$\mu$m thickness.
   The detector body is made of grade-2 titanium and supported against
an internal pressure by parallel and slant bars.
   The slant bar is designed to avoid complete obscuration of a
detector mask element.
   The geometrical area of the entrance windows is 88 cm$^{2}$.  Its
effective area as a function of the energy is shown in
figure~\ref{fig:pspc_effective_area} for vertical X-ray incidence.

\begin{figure}
   \centerline{\FigureFile(0.5\textwidth,){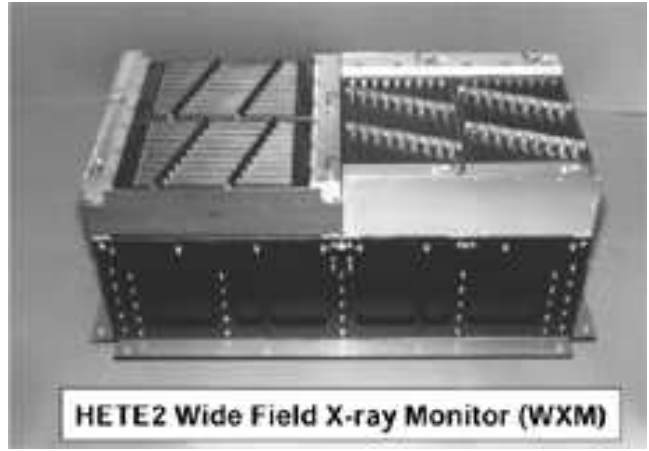}}
   \caption{Photograph of the WXM detector. 
            Four proportional counters are attached to the electronics
            box.
            The X detector is comprised of the left two counters.  It 
            measures the projection angle of the burst direction in the
            XZ plane.  The Y detector is comprised of the right two counters.  It 
            measures the projection angle in the YZ plane.
            The Z axis is along the vertical direction in the picture,
            while X axis points from right to left.
            The X detector consists of the XA (far side) and XB
            detectors (near side), while the Y detector consists of
            the YA (left) and YB (right) detectors.
            \label{fig:wxm_photo}}
   \label{fig:wxm_photo}
\end{figure}

\subsection{Electronics and Processing}

\begin{figure}
  \parbox{\textwidth}{
     \centerline{\FigureFile(0.8\textwidth,){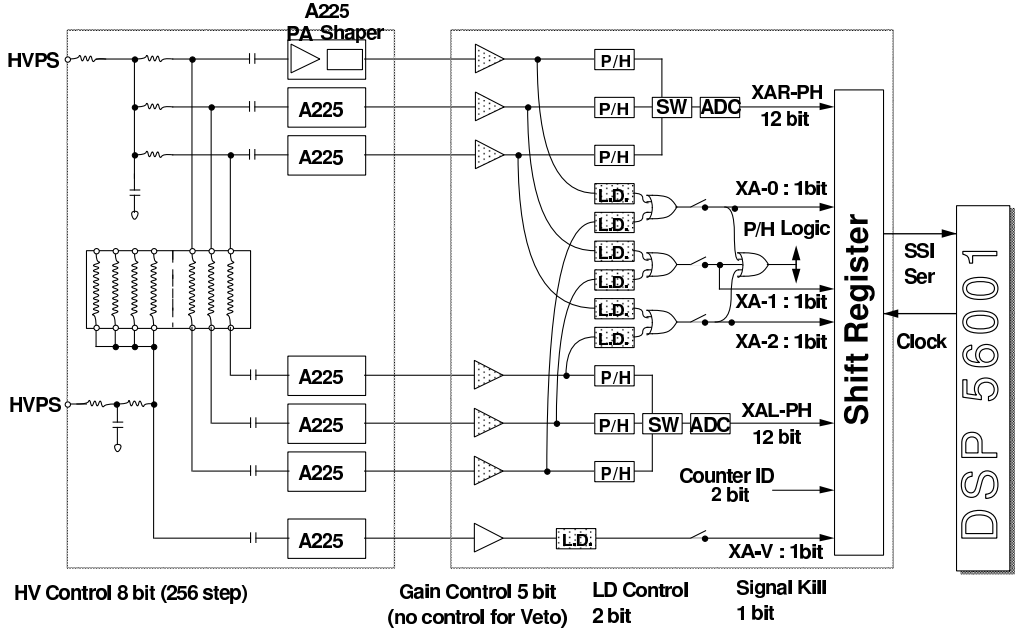}}
     \caption{WXM signal processing scheme for one counter.
              The same scheme applies to all four counters.
              Details are described in the text.
              \label{fig:wxm-e_block}
     }
  }
\end{figure}

   A block diagram of WXM signal processing is shown in
figure~\ref{fig:wxm-e_block}.
Analog
signals from both ends of the anodes are 
processed separately by charge-sensitive preamplifiers
(Amptek A225), while the signals from the four veto anodes are 
summed before the processing. 
   Seven A225s are used for each counter.
   The output of each A225 is connected to peak-hold (P/H) and 
lower-level discriminator (LD) circuits through a main amplifier.
   The gain of the main amplifier is adjustable to 32 levels
with a dynamic range of 10, except for the veto signal.
   The LD level is selectable from 4 levels of 100, 140, 200, 
500 AD units (which correspond to 0.4, 0.6, 0.9, and 2.5
keV, respectively), for a standard operation mode and for X-ray incidence
at the center of the anode.

   The logic for the AD conversion starts when any one of the
anode signals exceeds the lower discrimination level.
   AD conversion is performed only for the anode which has 
generated a LD signal.
   In the case of an LD hit for multiple anodes, one anode is 
selected by a pre-determined priority (wire~1 $>$ 0 $>$ 2).
   In this way, pulse heights induced at both the left and right sides
of an anode are digitized to 12-bit precision.
   Event data is expressed in two 16-bit words, comprising
2$\times$12-bit of ADU, 2-bit of counter ID, 4-bit of hit pattern flag
representing the LD hit for the 3 anodes and one veto, and two 1-bit 
word order flags.
   Background events caused by charged cosmic rays are identified as
multiple anodes/veto hit events, and can be rejected using the hit
pattern flag.

   The data loaded on the shift register is transmitted to the 
X-DSP after the previous data transmission is complete.
   In view of the fact that the data transfer time is 32~$\mu$s and signal processing
time  is $\sim$ 20~$\mu$s, we can estimate that even for a count rate of
12500~c s$^{-1}$ the dead time fraction is at most 10\%.
   If more than one detector is waiting to transfer data,
one of them is selected by a round-robin cyclic priority.
   This cyclic priority scheme assures that the data can be taken
uniformly for all of the detectors, even when one of the detectors has a much
higher count rate due to an anomalous condition such as electric discharge.

\subsection{CPU Processing}

   The spacecraft computer system consists of four identical processor 
boards: each board contains one transputer (INMOS T805), two Digital Signal
Processor (Motorola, DSP56001), and 20 Mbytes RAM. 
   The processors are assigned to the spacecraft and science nodes 
in the following way: 
  node 0: spacecraft processing,
  node 1: SXC,
  node 2: optical cameras, 
  node 3: WXM and Fregate.
   The ``links'' feature of the transputer allows for quick and efficient 
communications between processors. 
   The DSP serves as an interface to the instrument. 

   The WXM and FREGATE share a common processor board (XG node), and each
has a dedicated DSP (X-DSP and G-DSP) for handling the event data 
from the instruments, producing various data products and commanding 
the instruments.
   For the safety of the detector, the X-DSP provides a self-safety
mechanism to the WXM: an HV shutdown command is executed when the total
count rate exceeds 20000 counts per 4~s, or when the count rate in one
of the  four counters goes down below 5 counts per 0.8 s.

   The data products generated on the X-DSP and XG transputer are
summarized in Table~\ref{tbl:data-products}.
   Multiple-wire hit events are rejected prior to constructing those data
products, so the background rate is reduced in the products.
   The time history and the position histograms generated by the X-DSP
are transferred to the XG transputer.  The time history is used by the transient
event detection trigger, while the position histograms are used to localize
events detected by the trigger.
   The FREGATE time history is also transferred to the XG transputer
and used for triggering.
   These data are transferred to the telemetry after reducing the time
resolution by factors of 15 and 20 for time history and position histograms,
respectively. 

   The triggering time scales implemented as of 2003 May are from
80~ms to 27~s, and the threshold levels are set to 4.7--8.0 sigma,
depending on the time scale.
   To reduce the false trigger rate due to background variability, two
background regions are taken before and after the foreground region
and the expected background level in the foreground region is estimated
by the linear interpolation.
   This method, however, introduces a localization delay which ranges from
5~s  (trigger with 80~ms foreground) to 29~s (13~s foreground),
depending on the  triggering time scale.
   Thus, in the part of the orbit where the background rate is almost
stable, one-sided triggers (which take background only before the
foreground region) are used to reduce the delay time.
   The foreground time scales of the one-sided triggers are 5, 19 and 27~s.
   From the experience of recent in-orbit operation, the on-board
localization delay time is found to be 5.5~s to 43~s, as
summarized in table~\ref{tbl:WXM-GRBs}.
   Once a trigger is found, position histograms at the corresponding
foreground and background region are extracted from the memory, and the
localization procedure is applied to the data.
   When localization is completed, the trigger process continues to
search for better time intervals based on the signal-to-noise ratio, and
localization is performed for each updated foreground/background
region.
   The localization is performed by the cross-correlation method.
   The cross-correlation score, $C/\sigma_{C}$, is calculated as:
\begin{equation}
   C(j) = \psi_{j} [n \sum_{i} M_{ij} S_{i} - \sum_{i} S_{i} ],
   \label{eq:cc_c}
\end{equation}
\begin{equation}
   \sigma^2_{C(j)} = \psi_{j}^{2} [n^{2} \sum_{i} M_{ij}^{2} \sigma_{S_{i}}^{2} 
                   + \sum_{i} \sigma_{S_{i}}^{2} ],
   \label{eq:cc_sigma}
\end{equation}
   where subscript $i$ and $j$ represent position and directional bins,
respectively, $\psi$ is the normalization constant, $S$ is a
background-subtracted observed histogram, and $\sigma_{S}$ is the standard
deviation of $S$.
   $M$ is a ``template'', a numerical simulation of the detector hit
pattern for the corresponding direction.
   The burst direction is estimated by maximizing the cross-correlation
score over possible directions.
\begin{table}
   \begin{center}

   \caption{Data products generated by X-DSP and XG transputer.}
   \label{tbl:data-products}
   \begin{tabular}{lp{5em}p{5em}p{10em}p{0.43\textwidth}} \hline \hline
    Processor & Data type & Time resolution & Destination & 
                                    \centerline{Description} \\
    \hline
    X-DSP & HK  & 4 s & Telemetry &   
                                House Keeping (HK) data of the
                                instrument, such as power status, HV
                                setting, temperatures of an electronics
                                board and the support structure wall.
    \\ \cline{2-5}
    & TH  & 80 ms & XG Transputer &                              
                                Time histories for 4 energy bands
                                (approximately 2--5 keV, 5--10 keV,
                                10--17 keV and 17--25 keV) and for 
                                4 counters.
    \\ \cline{2-5}
    & POS & 330 ms & XG Transputer &
                                  Time-resolved position histograms
                                  for 2 energy bands (approximately 
                                  2--7 keV and 7--25
                                  keV) and for 4 counters.
    \\ \cline{2-5}
    & PHA & 4.9 s  & Telemetry &
                                  Time-resolved energy spectrum 
                                  for 4 energy bands (same as TH) and 
                                  for 12 anode wires.
    \\ \cline{2-5}
    & TAG & 256 $\mu$s  &         Telemetry & 
                                  Burst photon data tagged with
                                  time (8-bit), 
                                  position (7-bit),
                                  energy (5-bit), and
                                  wire ID (4-bit). This data type
                                  is generated in the burst observation
                                  mode.
    \\ \cline{2-5}
    & RAW & 1 $\mu$s  & Telemetry & 
                                  2 words of unprocessed
                                  raw photon data from the instrument and 2
                                  words of a time stamp. This data type
                                  is generated in the health check mode,
                                  3 minutes per day.
    \\ \hline
Transputer
    & TH & 1.2 s & Telemetry &     This is made by rebinning the
                                   time histories received from the X-DSP.

    \\ \cline{2-5} \cline{2-5}
    & POS & 6.6 s & Telemetry & 
                                  This is made by rebinning the position
                                  histograms received from the X-DSP.
    \\ \cline{2-5}
    & Burst Information 
    & Burst mode &  Telemetry, VHF and Trigger Monitor on another node &    
                                  Results of the trigger and
                                  localization processes are reported
                                  to the ground through S-band and VHF,
                                  and to the trigger monitor process for
                                  setting HETE-2 into the burst processing
                                  mode.
                                  \\ \hline
           \multicolumn{5}{c}
           {\parbox{180mm}{\footnotesize 
            Note. The ``destination'' column represents where the data is
            transferred. ``Telemetry'' means that the data is transmitted
            to the ground over the S-band radio.
            TH and POS data are transferred to the XG transputer for
            further processing of burst detection and localization,
            then transmitted to the ground after reducing the time
            resolution.}}
   \end{tabular}
   \end{center}
\end{table}

\section{Performance of the Proportional Counter}
\subsection{Experiment}

    To calibrate the response of the proportional counter (PC), we
irradiated a counter using characteristic X-rays of chlorine ($K_{\alpha} = 2.6$ keV),
titanium ($K_{\alpha} = 4.5$ keV), iron ($K_{\alpha} = 6.4$ keV), copper
($K_{\alpha} = 8.0$ keV), and molybdenum ($K_{\alpha} = 17.2$ keV) using
an X-ray generator on the ground, and measured the positional and
energy response.
    The X-rays were collimated to 0.2 mm diameter and irradiated 
at 127 $\times$ 54 grid points with 0.94 mm spacing.
    The chlorine characteristic X-ray was obtained by putting
a PVDC film of 20 $\mu$m thickness in front of the Be window
and irradiating it with titanium K X-rays, while the
others were obtained directly from the primary targets.
    For more uniform and wider range of energy calibration,
an experiment using a monochrometer was also performed for the 6--26~keV energy
range, with a 2 keV step.
    In these experiments, the bias voltage to the anode wire was changed from
1400~V to 1700~V to obtain the energy response for a wide range of gas gain.

   After installing the WXM on the spacecraft, positional and energy 
calibration was performed by using $^{55}$Fe radioisotopes attached at
the top of the support structure of the coded mask.
   The radioisotope is contained in a slit case and both sides of each
counter are irradiated by two radioisotopes at 40~mm distance from the
center of the counter.
   The detection count rate for each radioisotope is 1--2~c~s$^{-1}$.
    
\subsection{Positional Response}
\begin{figure}
   \parbox[t]{0.49\textwidth}{
      \centerline{\FigureFile(0.43\textwidth,){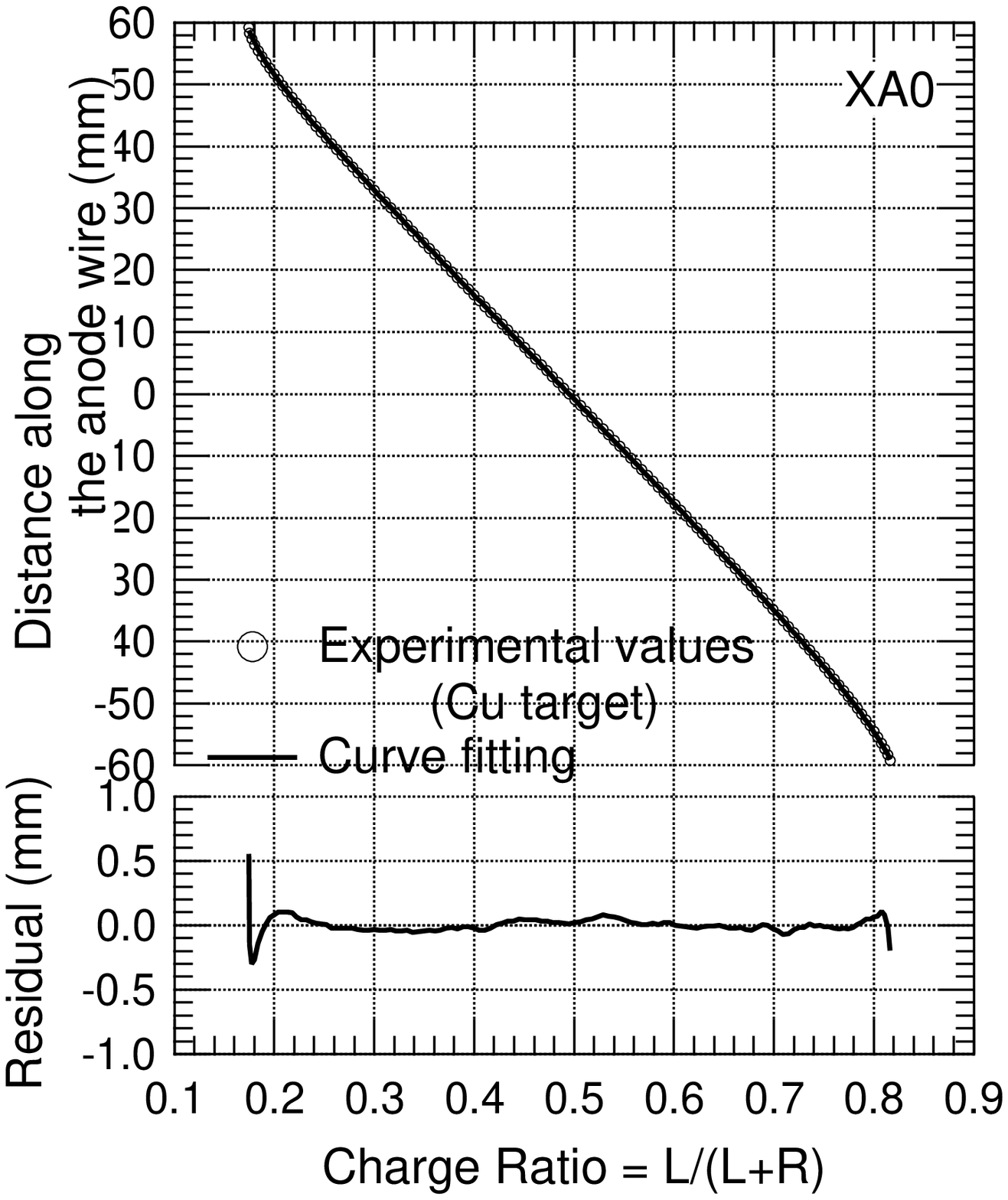}}
      \caption{Relation between the actual X-ray absorption position
               and the position measure defined by
               $L/(L+R)$, where $L$ and $R$ represent the pulse heights
               measured at the left and right sides, respectively.}
      \label{fig:pm-x}
   }
   \parbox[t]{0.49\textwidth}{
      \centerline{\FigureFile(0.43\textwidth,){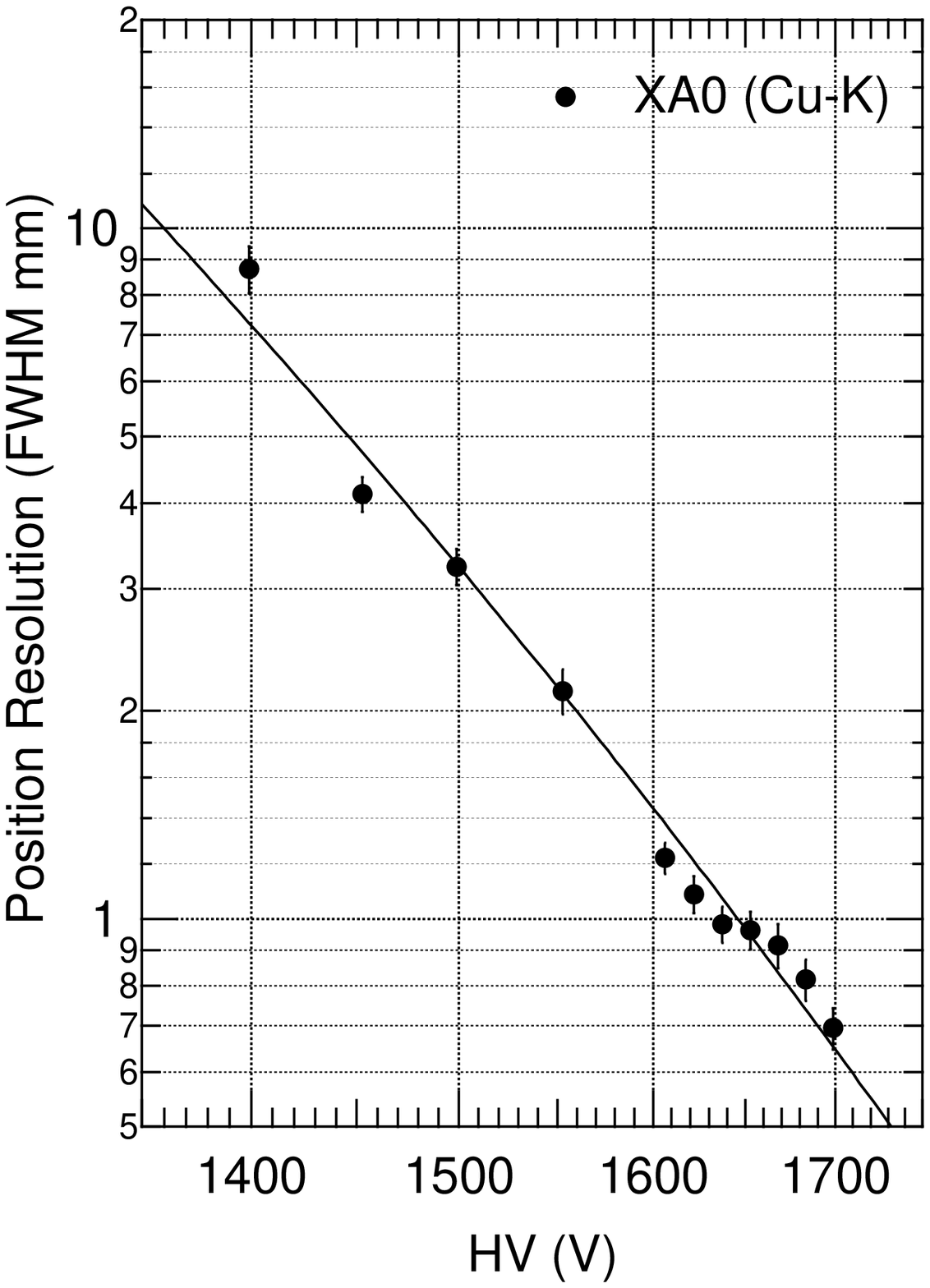}}
      \caption{Positional resolution vs. bias voltages for 8 keV X-rays.}
      \label{fig:pres_vs_hv}
   }
\end{figure}

    We have measured a relation between the Position Measure,
$PM = L/(L+R)$ ($L$: pulse height measured at the left side of anode,
$R$: pulse height at the right side), and the X-ray absorption
position measured along the anode wire (figure~\ref{fig:pm-x}).
    To derive the position measure at each position $X$, we stacked
the data obtained by scanning perpendicularly across the anode wire 
at each fixed $X$-position, and calculated the median $PM$.
    The resulting relation deviates from linearity at both ends of
the anode wire, mainly because of electric field distortion.  The relation
can be expressed by the following empirical formula to within 0.2~mm error:
\begin{eqnarray}
  \label{eq:s-curve}
  X \hspace{0.1cm} \mbox{(mm)} 
                & = & a_{1} + a_{2} \times PM + \frac{a_{3}}{PM - a_{4}} - \Delta x
                                    \hspace{1cm} PM \ge a_{0}, \\
                & = & a_{5} + a_{6} \times PM + \frac{a_{7}}{PM - a_{8}} - \Delta x
                                    \hspace{1cm} PM <  a_{0}, \nonumber
\end{eqnarray}
    The coefficients $a_{0}$--$a_{8}$ were obtained independently for
each anode.
    We noticed that the PM--X relation is gradually changing, as described
in subsection~\ref{sec:flight_pos_cal}, so these parameters are
updated every six months based on in-flight calibration.
    The parameter $\Delta x$ is a small correction term which is
determined using the in-flight calibration.
    This systematic offset represents the effect of mechanical tolerance
of the experimental alignment on the ground calibration.

    Fig.~\ref{fig:pres_vs_hv} shows the positional resolution for Cu-K
X-rays at each bias voltage.
    The primary purpose of the WXM is to localize GRBs with
$\sim$~10$'$ accuracy.
    This can be achieved with a positional accuracy of less than 1~mm
for a 187~mm focal length, so the bias voltage should be greater than
1650~V.
    The positional resolution is mainly limited by Johnson
noise, due to the relatively low resistivity of the carbon fiber.
    The positional resolution measured in this calibration experiment is
expressed by the sum in quadrature of three components.
    The dominant component is due to electronic noise.
    The other contributions are statistical fluctuations of the
center of electron clouds, and the finite X-ray beam size ($\sim$ 0.2~mm).
    Since the amplitude of the Johnson noise is inversely proportional
to the square root of the resistance, it is expected that the positional
resolution is approximately a function of $(L+R) \times \sqrt{R_{i}}$.
    $R_{i}$ is the resistance of the $i$-th anode wire, which ranges from 12.0 
to 15.8~k$\Omega$.
    As demonstrated in \citet{wxm.SPIE.2000}, the resolution
can be well expressed as a function of the signal-to-noise ratio.
\begin{figure}
   \centerline{\FigureFile(0.9\textwidth,){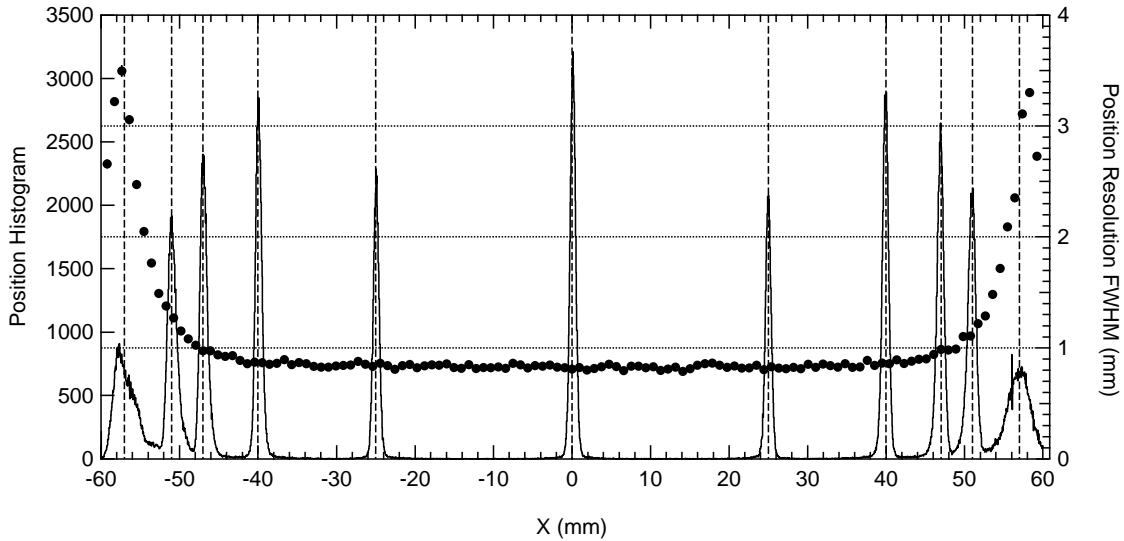}}
   \caption{Positional dependence of the position resolution.
            The histogram shows the position distribution for X-ray
            incidence at 11 X-positions shown with dashed lines, 
            and the solid circles show the positional resolution 
            measured in FWHM of the distribution.}
   \label{fig:pos_res_vs_x}
\end{figure}

    Since the direction of the electric field at the edges of the
counter is not perpendicular to the direction of anode wire due to the
internal structure, the X-position of gas amplification region systematically
differs from that of X-ray absorption according to the three-dimensional 
absorption position.
    Thus, it is expected that the positional resolution is degraded at the
counter edges.
    Figure~\ref{fig:pos_res_vs_x} shows the error distributions at eleven
$X$-positions (histograms) and positional resolution measured in FWHM
(solid circles).
    They are obtained by accumulating calibration data at fixed
X-positions of XA1 anode.
    The resolution is almost constant in the central region (from 
$-$40~mm to $+$40~mm), but it is substantially degraded as the position
approaches the edge of the detector, due to the distortion of the electric
field.

\subsection{Energy Response}
\begin{figure}
   \centerline{\FigureFile(0.8\textwidth,){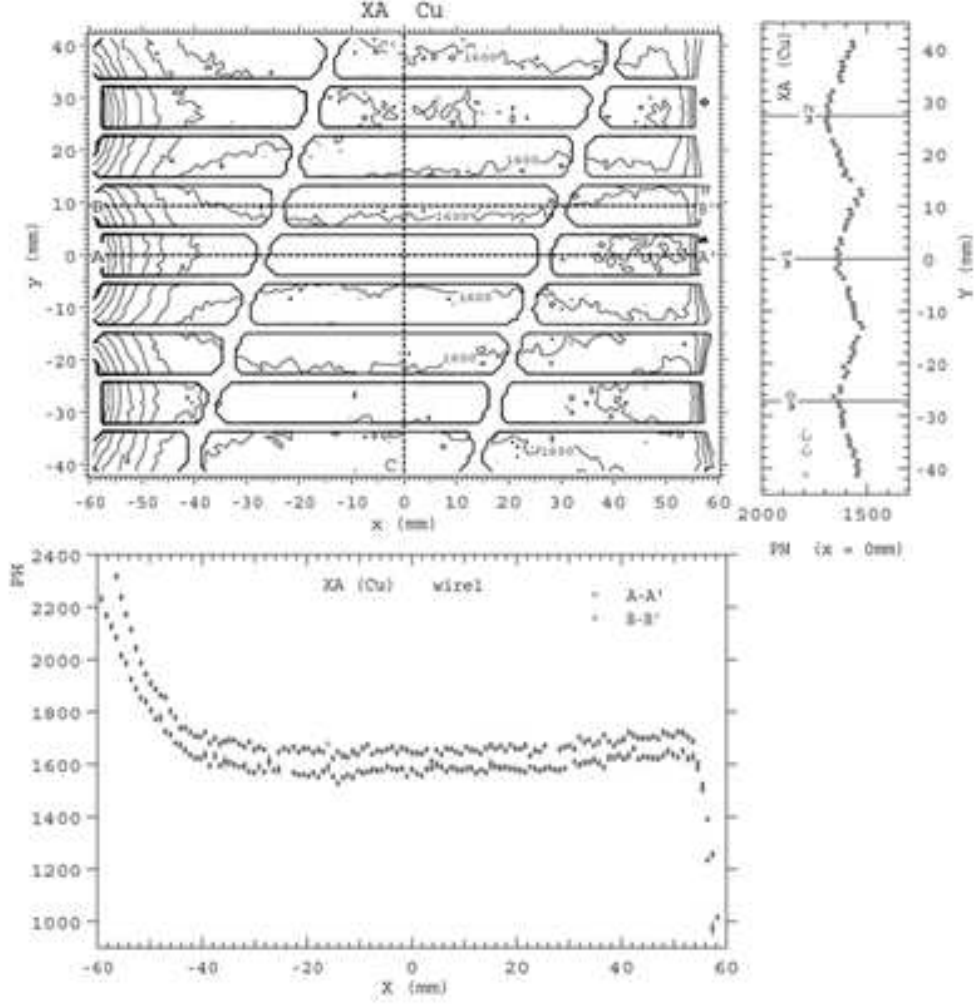}}
   \caption{Gain uniformity of the XA detector at Cu K energy.
            The left-top panel shows a contour plot of the gain
            distribution, the bottom panel shows the gain distribution
            along the line of A--A$'$ and B--B$'$, and right panel shows
            the gain distribution along the line of C--C$'$.}
   \label{fig:ph_2d-map}
\end{figure}

   In figure~\ref{fig:ph_2d-map}, a 2-dimensional distribution of the average
PH for the Cu-K X-ray (K$_{\alpha}$ = 8.04~keV) is shown together with the 
1-dimensional projections at sections  A--A$'$, B--B$'$, and C--C$'$.
   One can see that the gas gain increases at the left side, while
it decreases at the right side.
   This is due to the internal mechanical structure of the PC.

\begin{figure}
   \centerline{\FigureFile(0.8\textwidth,){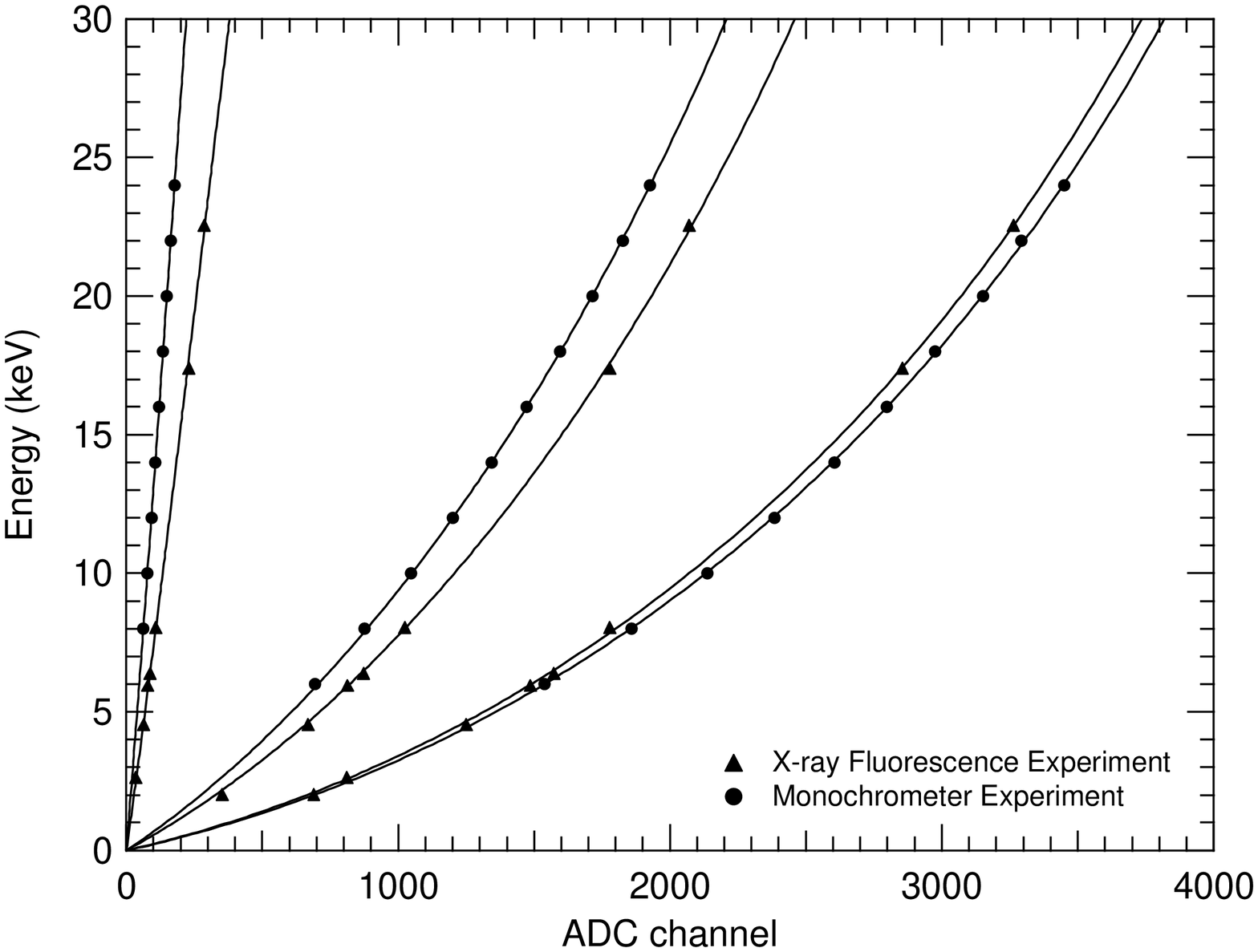}}
   \caption{X-ray incident energy vs. measured pulse height. The solid
   circles show the experimental data and the lines show the result of
   parameterization using  equation~(\protect\ref{eq:ph2ene}), the
   coefficients of which were derived as a function of equivalent bias voltage 
   parameter $v$.
   From left to right, the curves correspond to voltages $v =$ 1299, 1361, 1599, 1618, 1694, 1699~V, 
   respectively.}
   \label{fig:ene_vs_ph}
\end{figure}

   In figure~\ref{fig:ene_vs_ph}, the relations of the PH to the X-ray 
incident energy for various bias voltages (1300--1700~V) are shown
for XA0 anode.
   The X-rays are irradiated at the position offset by 10~mm in the 
direction perpendicular to the wire.
   As shown in this figure, the PH is not linearly related to the
energy at voltages above 1600~V, because of a self-induced space-charge
effect.
   As mentioned in the previous section, we operate the PCs  around
1650~V.  We have therefore found it necessary to obtain a non-linear
formula to translate PH to energy.
   The experimental result is well represented by the following 
formula:
\begin{equation}
   E = a \cdot \left( \mbox{PH}/{c} \right)
       + b \cdot \sinh{ \left( \mbox{PH}/{c} \right)}
                                 + d \cdot \left( \mbox{PH}/{c} \right)^{2}
   \label{eq:ph2ene}
\end{equation}
   We assume that the relation depends only on gas gain and parameterize
the coefficients $a$, $b$, $c$, and $d$ as a function of bias voltage $v$
for the case of reference anode XA0.
   The gas gain is slightly different for each anode because of the
difference in the wire radius and electric field, so the equivalent
potential $v$ of the above formula is separately determined for each anode.
   The spatial non-uniformity of the gas gain is corrected by
introducing a spatial dependence to $v$.

\begin{figure}
  \centerline{\FigureFile(0.8\textwidth,){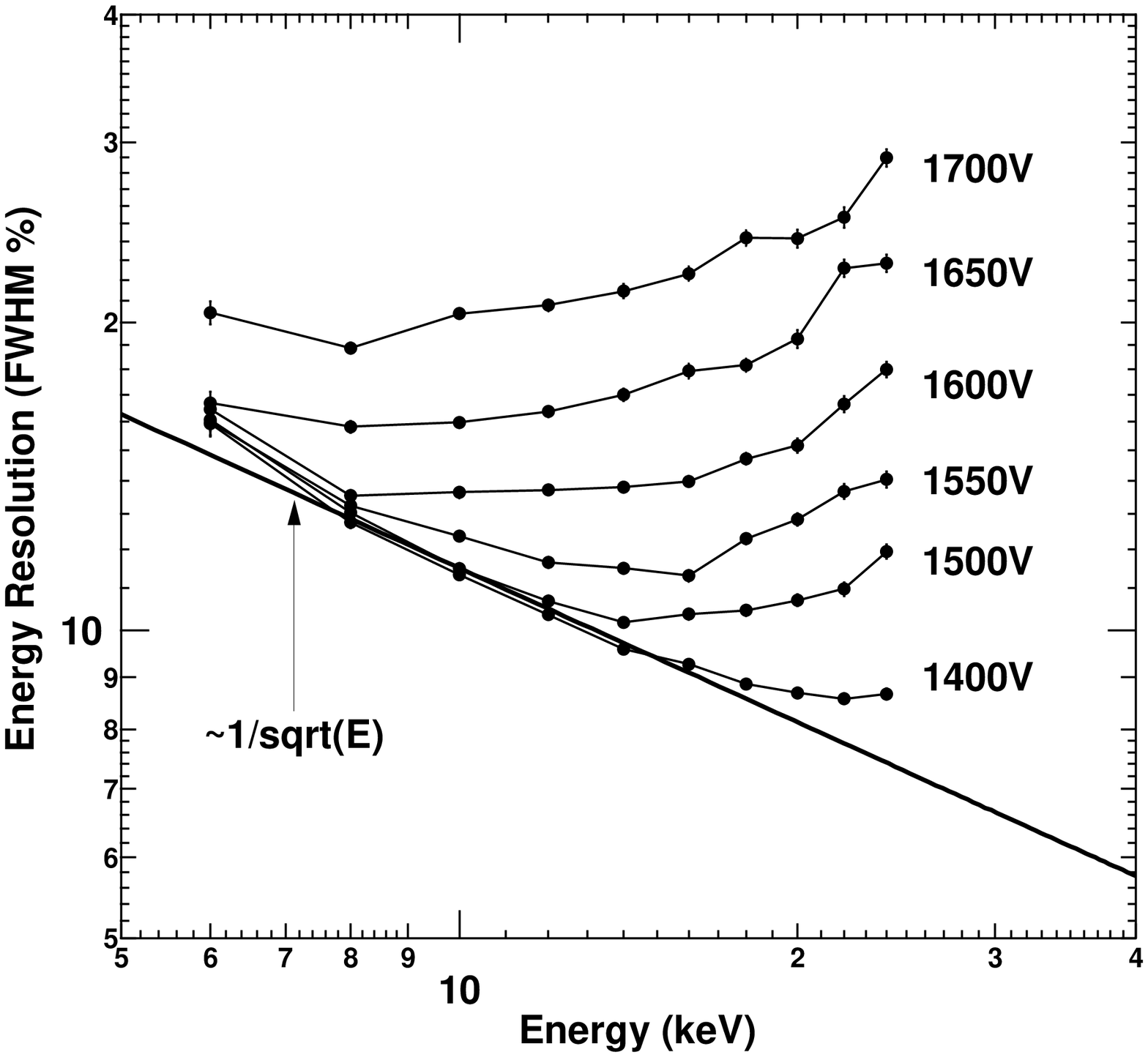}}
  \caption{Energy resolution at energies from 6 to 24 keV for bias
           voltages from 1400 to 1700 V}
  \label{fig:eres}
\end{figure}

   Using the PH-energy relation formula obtained above, the energy
resolution is determined for energies of 6--24~keV and for bias
voltages of 1400--1700~V.
   In figure~\ref{fig:eres} the results for off-wire incidence are plotted
against energy.
   However, at a low bias voltage of 1400~V, energy dependence of the
resolution satisfies the $1/\sqrt{E}$ law; at higher voltages it no
longer follows the law, owing to the space-charge effect.
   We find that the resolution is well expressed as a function of the
empirical parameter $M^{0.23} \times n_{0}$ for various bias voltages.
   Details are described in \citet{wxm.SPIE.2000}.

\section{RIKEN Ground Localization Method}
\label{sec:riken_method}

    A ground analysis of the GRB localization by the WXM is performed by
two independent methods, the Chicago Bayesian method~\citep{wh.2002.carlo} 
and RIKEN cross-correlation method~\citep{wh.2002.yuji}.
    The RIKEN localization method is based on the variance-weighted 
cross correlation method, which searches for the direction where the simulated
position histogram best matches the observed one.
    The procedure consists of two parts.
    First we determine a coarse location ($\theta_{x,0}, \theta_{y,0}$)
using the model histograms simulated for the directions spaced by 
0$^{\circ}$.2 along  $\theta_x = 0$ and $\theta_y = 0$ axes.
    Next the fine location is determined using models simulated
along $\theta_x = \theta_{x,0}$ and $\theta_y = \theta_{y,0}$
by every 0$^{\circ}$.05 step.
    In the model calculation, we take into account the following 
effects:
  (1) the slant penetration effect,
  (2) positional resolution, $\sigma(E,x)$, as a function of energy 
      and X-ray absorption position,
  (3) obscuration function, $f(x)$, by the PC body structure, the 
      side wall, and the coded mask.
   The model histogram for an X-ray of energy $E$ is expressed as
\begin{eqnarray}
   P(x,E) & = & \epsilon(E) \cdot  \frac{1}{\sqrt{2\pi\sigma(E,x')}} 
                \cdot \frac{1}{\lambda_{x}(E)} 
                \int_{x_{\mbox{\tiny min}}}^{x_{\mbox{\tiny max}}}
                dx'' f(x'') \nonumber, \\
& \times &   \int_{x''}^{x''+d \tan(\theta_x)}
       dx' e^{-\frac{(x-x')^2}{2\sigma^{2}(E,x')}} 
       e^{-\frac{(x'-x'')}{\lambda_{x}(E)}},
\end{eqnarray}
where 
$\epsilon$ is the transmission coefficient of the 100~$\mu$m Be window and
the 7.62~$\mu$m thermal shield,
$\lambda_{x}$ is the projection of the mean free path onto the wire direction, 
$x_{\mbox{\scriptsize min}}$ and $x_{\mbox{\scriptsize max}}$ are the
available range of anode wire, which is 
from $-$57 mm to $+$57 mm, and $d$ is the depth of an anode cell.
   Denoting the source spectrum by $F(E)$, the spectrum-weighted model
histogram is given by
\begin{equation}
   M(x) = \int F(E) P(x,E).
\end{equation}
   We usually assume an $E^{-1.5}$ spectrum for a GRB analysis.  For the
purpose of WXM astrometric calibration, we use an appropriate spectrum type for
each X-ray source.
   The cross-correlation score is calculated by equations~(\ref{eq:cc_c}) and
(\ref{eq:cc_sigma}).
   The score is evaluated at every 0$^{\circ}$.05, and the maximum value is
obtained with 0$^{\circ}$.01 accuracy by interpolation.

   The localization error is estimated by simulating 1000
sets of position histograms based on the observational data.
   The simulation is performed by introducing a Poisson fluctuation to the
observed position histograms for the foreground and background regions, and 
applying the same localization procedure as that applied to the observational
data.
A 90\% C.L. error circle and the 90\% C.L. rectangle are then calculated by
requiring that 90\% of the simulated locations are contained in the error
region.
   We choose a circle or rectangle depending on which of the two regions subtends the
smallest solid angle.
   The rectangular regions are usually preferable when the signal-to-noise
ratio in one detector is significantly smaller than in the other, e.g.
when a GRB occurs at the edge of the FOV in the X direction but not
in the Y direction.

\section{In-Orbit Performance of the WXM}

   \subsection{Background}

   The WXM background consists of a non-X-ray background (NXB; charged
particles) which dominates in a restricted region of the HETE-2
orbit, the cosmic X-ray background (CXB) which dominates over the majority
of the orbit, and the steady X-ray point source background from sources
such as the Crab Nebula and Sco X-1.
   To reduce the NXB, the anti-coincidence method is used.
   In laboratory calibration, the total background count rate was
measured as 30 c~s$^{-1}$ for one counter, and after the anti-coincidence
and energy selection the background rate was reduced to 4.4 c~s$^{-1}$ in 
the energy range 2--25 keV, which corresponds to $4.7 \times 10^{-4}$
c~s$^{-1}$~cm$^{-2}$~keV$^{-1}$.

\begin{figure}
   \centerline{\FigureFile(0.8\textwidth,){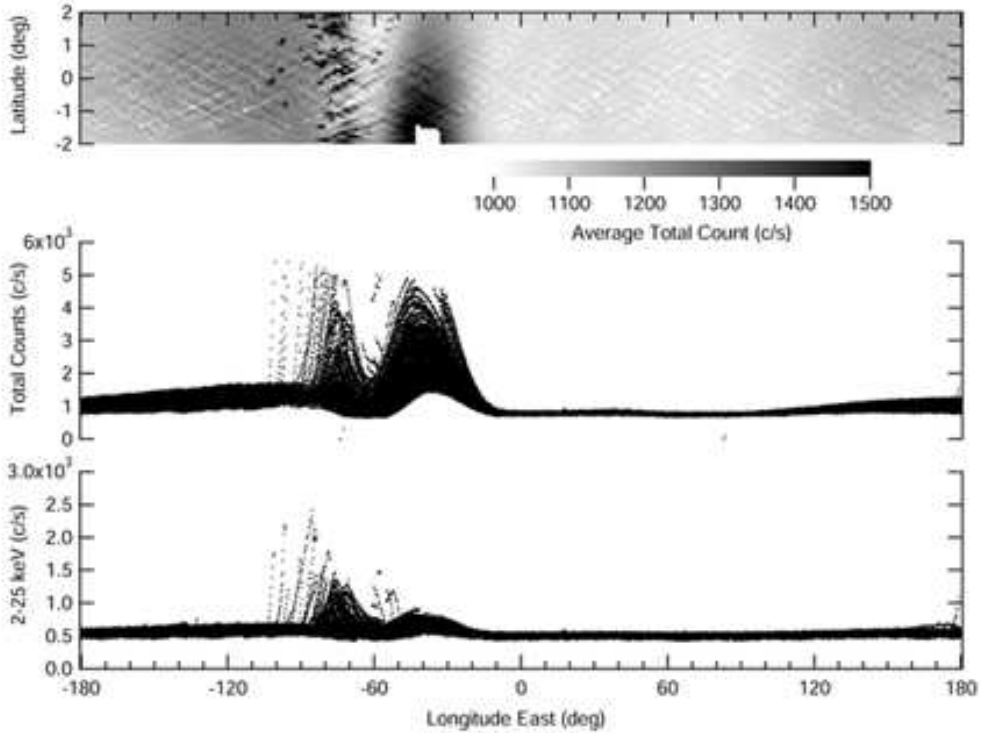}}
   \caption{Background counting rate in the HETE orbit.
     The top panel shows the averaged total counting rate at latitude
     from 
     $-2^{\circ}$ to $+2 ^{\circ}$. The blank at the center of SAA region is
     due to lack of observations at that longitude. The middle panel  shows a
     typical  total count rate distribution in one month. The bottom panel
     shows a typical count rate distribution of anti-coincidence events of
     2--25 keV in one month.
   }
   \label{fig:background}
\end{figure}

   The background count rate in orbit is mostly dependent on the
geographic coordinates of the spacecraft.
   The total count rate and anti-coincidence rate for the 2--25 keV
energy range are shown in figure~\ref{fig:background} as a function of 
the geographic coordinates.
   According to the background level and its time variability, the HETE-2
orbit is divided into 4 types of geographic regions:
\begin{enumerate}
   \item The most quiet region of longitude from $10^{\circ}$W to
	 $120^{\circ}$E over the African Continent, the Indian
         Ocean, and the Oceanian region,
         where the total count rate is very stable and as low as 1000
         c~s$^{-1}$. 
         In this region, the 2--25 keV event rate with anticoincidence
         is 530 c~s$^{-1}$, while the event rate without anticoincidence is 560
         c~s$^{-1}$.
   \item The moderate background region from $120^{\circ}$E to $105^{\circ}$W
         over the Pacific Ocean, where the total count rate is a time 
         variable in the range 1000--1500 c~s$^{-1}$.
   \item The pre-SAA region from $105^{\circ}$W to $60^{\circ}$W over the
         East Pacific Ocean and the Republic of Colombia, where the total
         count rate is usually the same as that of region 2, but can 
         drastically increase at times of strong solar activity.
   \item The SAA region from $60^{\circ}$W to $10^{\circ}$W over Brazil and
         the West Atlantic Ocean, where the count rate exceeds
         5000 c~s$^{-1}$.
\end{enumerate}

   By applying the anti-coincidence condition, the proton background in
the SAA region is effectively reduced by more than 90\%, but the transient
events in the pre-SAA region are not effectively rejected.
   This is because the transient events mostly consist of
electrons and/or positrons which have a shorter mean free path in the
proportional counter than the dimensions of an anode cell, so they are
absorbed in a single cell.

The measured counting rates for 2--25 keV anti-coincidence events due to the Crab
Nebula and Sco X-1 are 270 c~s$^{-1}$ and 2040 c~s$^{-1}$, respectively, when they
are at the on-axis direction.

\begin{figure}
   \centerline{\FigureFile(0.8\textwidth,){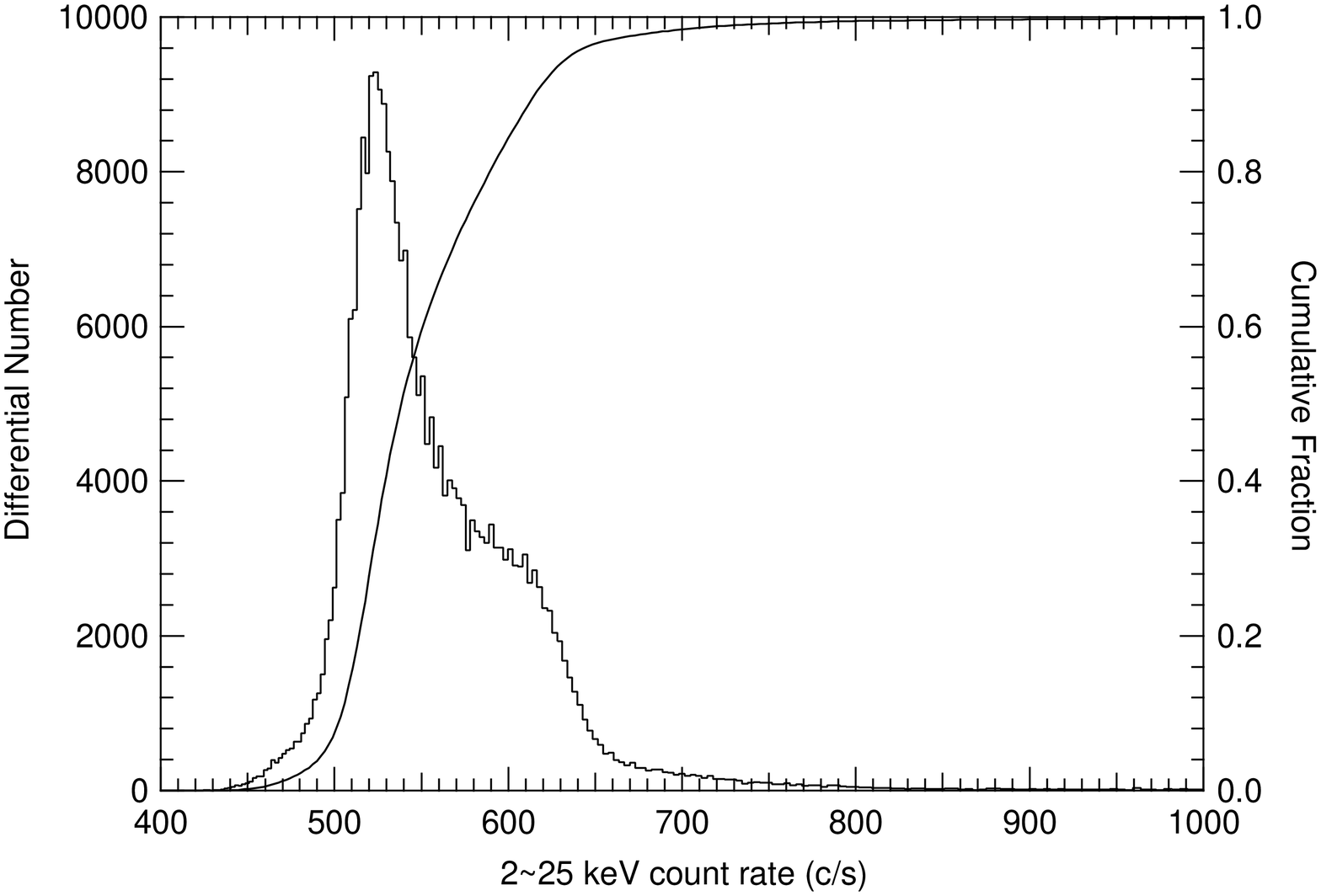}}
   \caption{Background count rate distribution at the 2--25 keV energy
            range. The histogram shows the differential distribution and
            the solid line shows the cumulative distribution.}
   \label{fig:bg_dist}
\end{figure}

In figure~\ref{fig:bg_dist}, the probability distribution of the
background rate in 2001 October is shown for 2--25 keV
anti-coincidence events.
The mode of the background rate is $\sim$ 530 and in 96\% of the WXM 
on-time the background rate is less than 640 c~s$^{-1}$.
Assuming a GRB of 10~s duration and a background rate of 530 c~s$^{-1}$,
the 5-sigma trigger threshold corresponds to 0.13 Crab.

   \subsection{In-Orbit Calibration of Positional Response}
\label{sec:flight_pos_cal}

\begin{figure}
      \centerline{\FigureFile(0.8\textwidth,){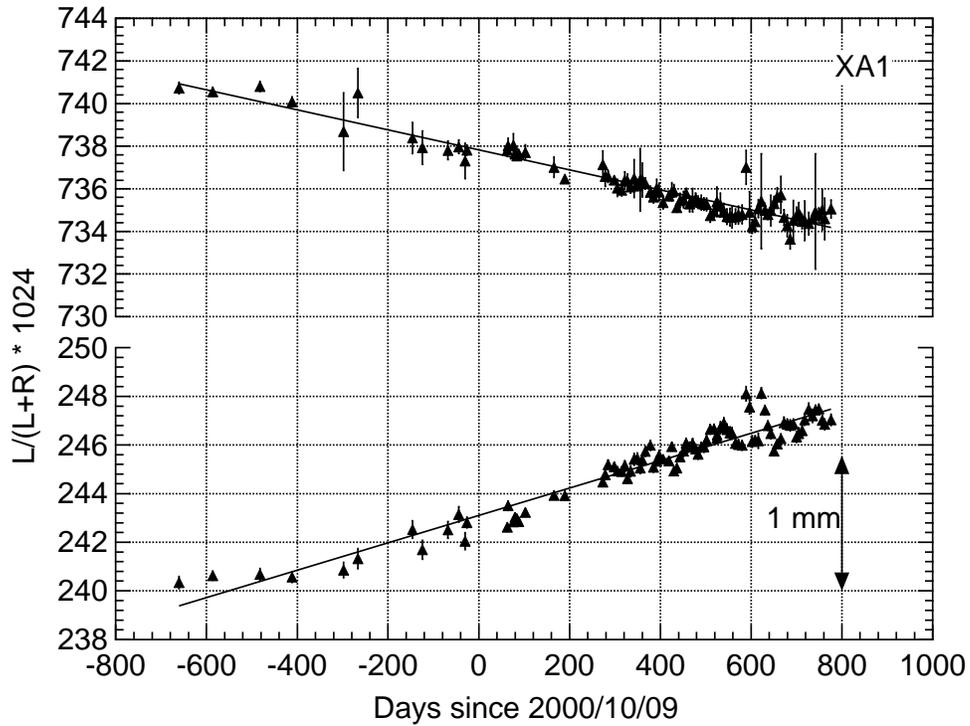}}
      \caption{Long-term variation of the position measures of the
               calibration sources for the case of an XA0 wire.
               Day 0 corresponds to the HETE-2 launch date.}
      \label{fig:cal_source_pm_variation}
\end{figure}

    From long-term monitoring of the signal produced by the
radioisotopes, we noticed that the apparent position calculated by 
equation~(\ref{eq:s-curve}) is gradually changing at a rate of 1 mm per 3 years
at the position irradiated by the calibration sources.
    The variation of the position measured for XA1 is shown in
figure~\ref{fig:cal_source_pm_variation}.
    The cause of this variation is not understood yet, but is likely due
to a change in the metallic contact of the carbon wire.
    In order to correct this variation, we assumed that the correction
is linear in the position measure.

    After the launch of the spacecraft, we performed a consistency 
check of the positional response among all of the anodes by localizing Sco~X-1
using each individual anode.
   The average difference for each anode is summarized in
table~\ref{tbl:offset}.  The corresponding positional offset defined in
equation~(\ref{eq:s-curve}) is also listed there. 
   We interpret these systematic errors as the effect of the mechanical tolerance
of the experimental alignment on the ground calibration.
We correct them
by subtracting the offset in the equation~(\ref{eq:s-curve}).
\begin{table}[hbt]
   \begin{center}
   \caption{Systematic difference of the calculated directions of Sco  X-1.}
   \label{tbl:offset}
      \begin{tabular}{ccccc} \hline
          & $\langle \Delta\theta \rangle$ &  $\sigma$ & $N$   &  $\Delta x$  \\
          &  (deg)    &  (deg)    &            &   (mm)    \\ \hline \hline
      XA0 & $-$0.010  &   0.028   &   43       &  0.034    \\
      XA1 &           &           &            &           \\
      XA2 &  0.066    &   0.025   &   41       & $-$0.216  \\
      XB0 & $-$0.072  &   0.024   &   40       &  0.235    \\
      XB1 & $-$0.016  &   0.019   &   44       &  0.052    \\
      XB2 &  0.002    &   0.035   &   40       & $-$0.005  \\ \hline
      \end{tabular}
      \hspace{1em}
      \begin{tabular}{ccccc} \hline
          & $\langle \Delta\theta \rangle$ &  $\sigma$ & $N$   &  $\Delta x$  \\
          &  (deg)    &  (deg)    &            &   (mm)    \\ \hline \hline
      YA0 & $-$0.033  &   0.022   &   29       &  0.108    \\
      YA1 &           &           &            &           \\
      YA2 &  0.051    &   0.014   &   33       & $-$0.167  \\
      YB0 & $-$0.071  &   0.017   &   25       &  0.232    \\
      YB1 & $-$0.049  &   0.018   &   21       &  0.160    \\
      YB2 &  0.004    &   0.012   &   20       & $-$0.012  \\ \hline
      \end{tabular}
   \end{center}
      {\footnotesize Note.
           $\langle \Delta\theta \rangle$, average difference from the direction 
           obtained for reference anodes XA1 or YA1;
           $\sigma$, standard deviation of the difference from the
           average; $N$, number of samples used for this analysis;
           $\Delta x$, corresponding offset parameter
           of equation~(\protect\ref{eq:s-curve}).}
\end{table}

\begin{figure}    
   \parbox{0.5\textwidth}{
      \centerline{\FigureFile(0.5\textwidth,){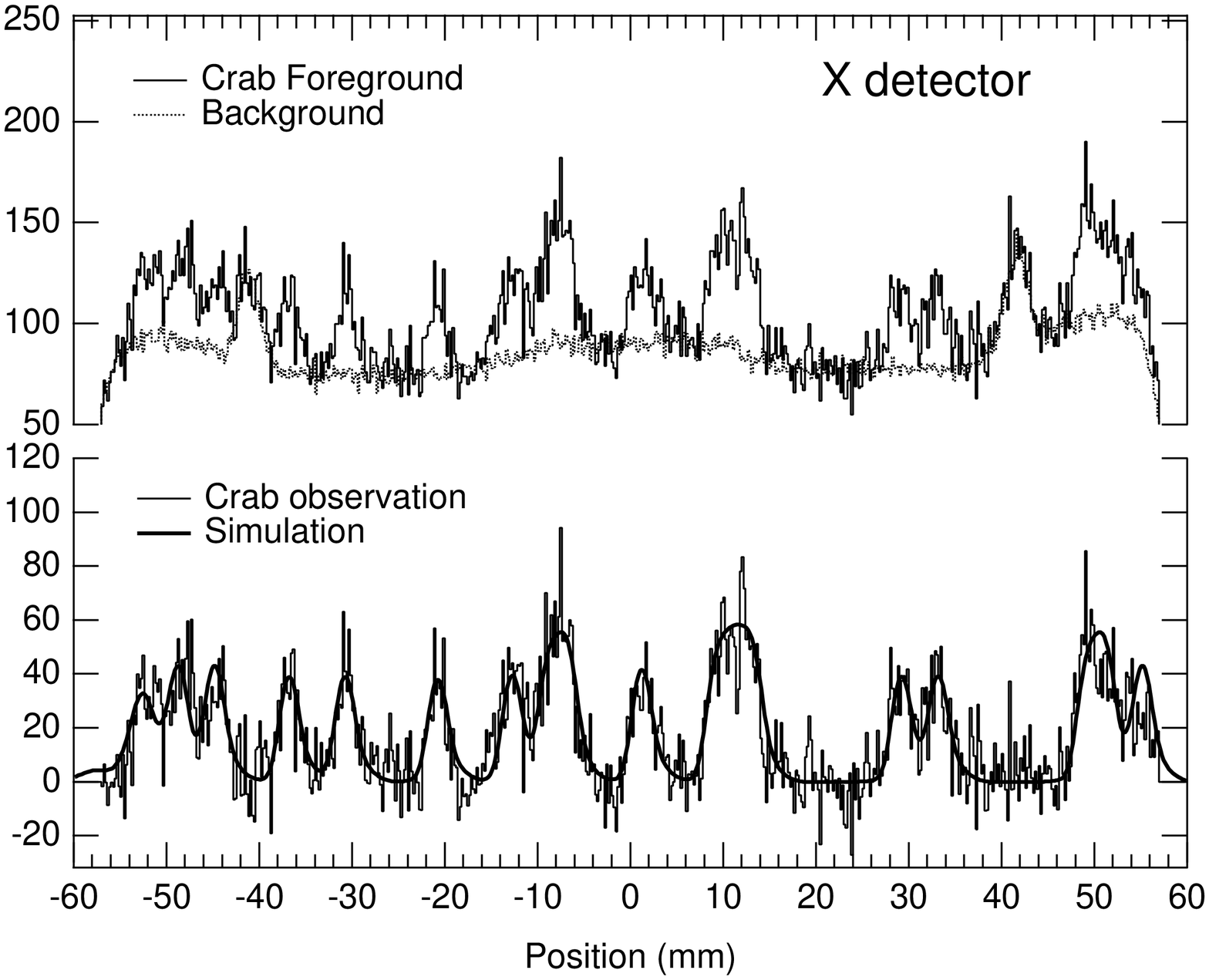}}
   }
   \parbox{0.5\textwidth}{
      \centerline{\FigureFile(0.5\textwidth,){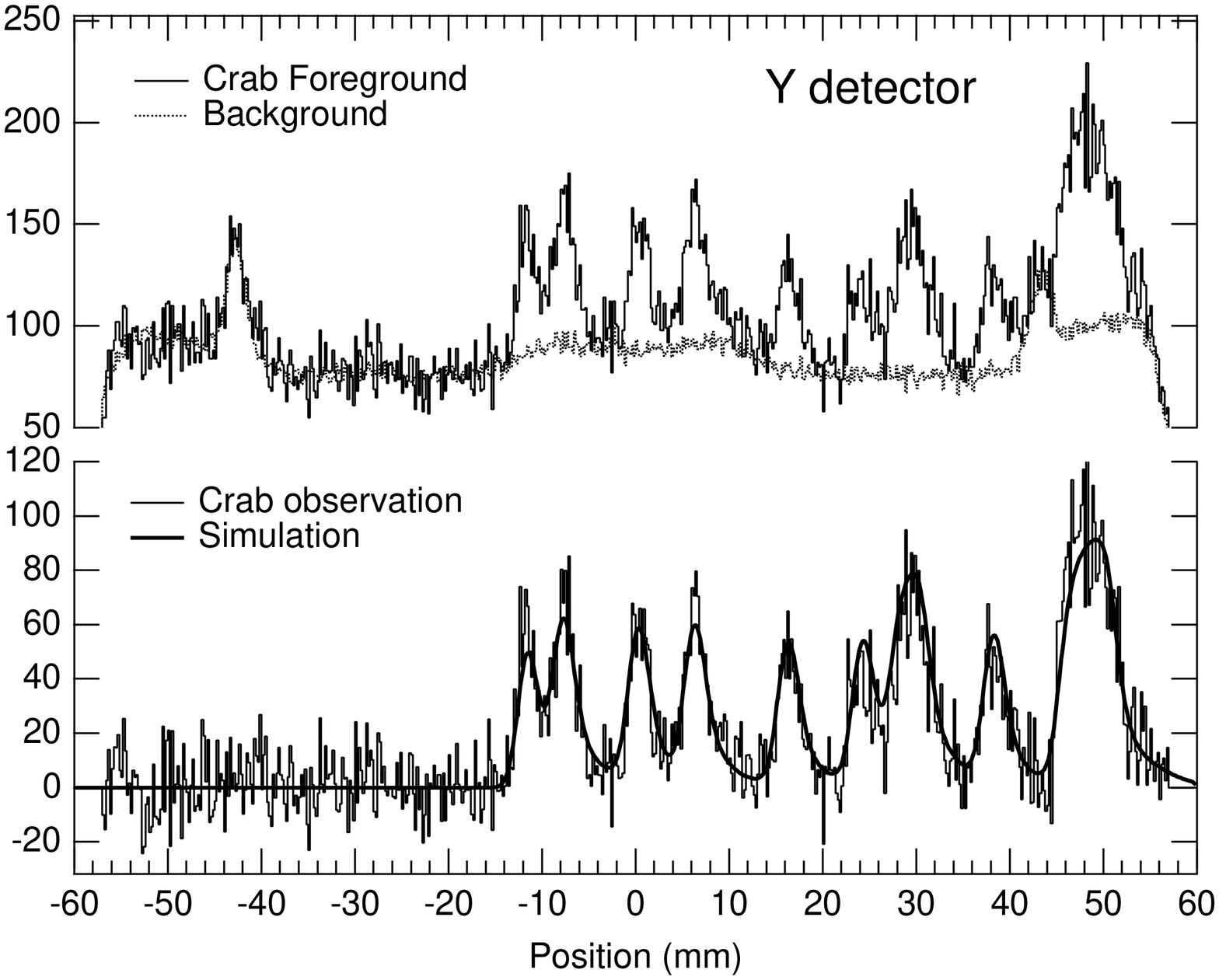}}
   }
   \caption{Crab Nebula images taken by the WXM. Each top panel shows
position histograms of the photons observed during the period
when Crab is in the FOV (solid line) and outside the FOV (dotted line).
Each bottom panel shows a comparison of background-subtracted Crab Nebula
images (histogram) and simulated images (thick line). The two peaks 
in the background histogram around $\pm$ 40 mm are due to the on-board
calibration isotopes.}
   \label{fig:crab_image}
\end{figure}

    Modeling the positional response, we simulated a mask
pattern measured by the WXM, and compared it with the observed Crab
Nebula image.
    The result is shown in figure~\ref{fig:crab_image}.
    The simulated Crab Nebula images are in good agreement with the
observed images.

\subsection{In-Orbit Gain and Spectrum Calibration}
   \label{sec:spect_crab}
\begin{figure}
   \centerline{\FigureFile(0.6\textwidth,){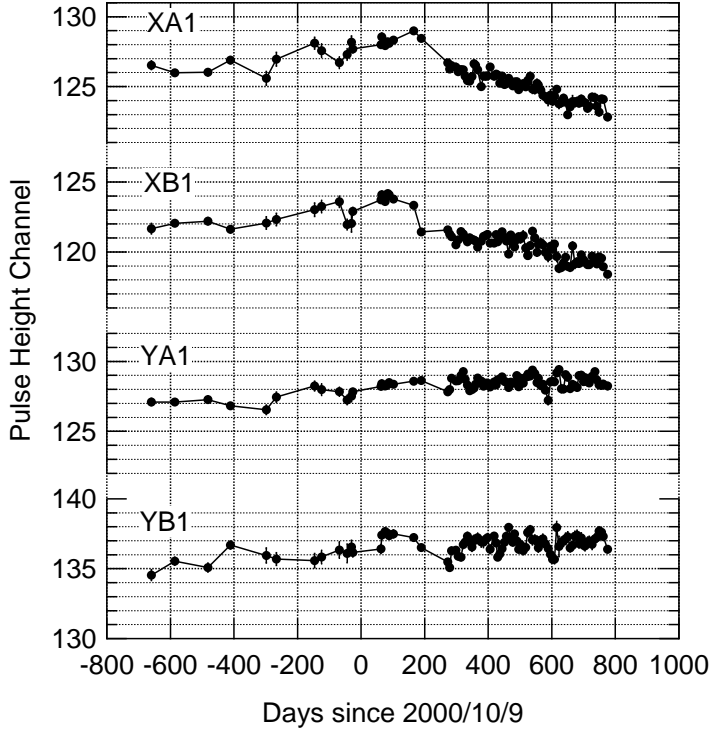}}
   \caption{Long-term gas gain variations monitored using the radioisotopes.
            From the top, XA1, XB1, YA1, and YB1. Day 0 corresponds to
            the HETE-2 launch date.
   }
   \label{fig:gain_history}
\end{figure}

   In figure~\ref{fig:gain_history}, long-term gas gain variations
monitored using the radioisotopes are shown for the XA1, XB1, YA1, and YB1
anodes.
   For the XA and XB counters, the gas gain gradually increased until
day 200 since the launch date, and decreased thereafter.
   For the YA and YB counters, the gas gain was still continuing to
increase at day 800.
   The calibration experiment was carried out at around day $-700$, so 
the maximum gain deviation from the ground experiment is about 2\%
as of day 800.
   The methods for calibrating the WXM energy response are limited to the
observations of Crab Nebula and monitoring the radioisotope signal,
since no other sufficiently bright celestial X-ray source is ever in the
WXM FOV, with the exception of Sco X-1.
   Because Sco X-1 is a strongly variable source, it is not suitable for 
an energy calibration.
   Using the Crab Nebula and the signal from the radioisotopes, we may
calibrate the temporal gain variation, the absolute gain scale, and the
detector efficiency.
   The detector efficiency and gain scale are checked using Crab Nebula
observations, and the temporal gain variation is corrected by monitoring
the gain using the signal from the radioisotopes, as described above.
   The other items not calibrated in-orbit, such as the energy
resolution, the spatial gain variation, and the energy pulse height linearity,
rely on the ground calibration experiment.

\begin{figure}
   \centerline{
      \parbox[t]{0.5\textwidth}{
         \FigureFile(0.5\textwidth,){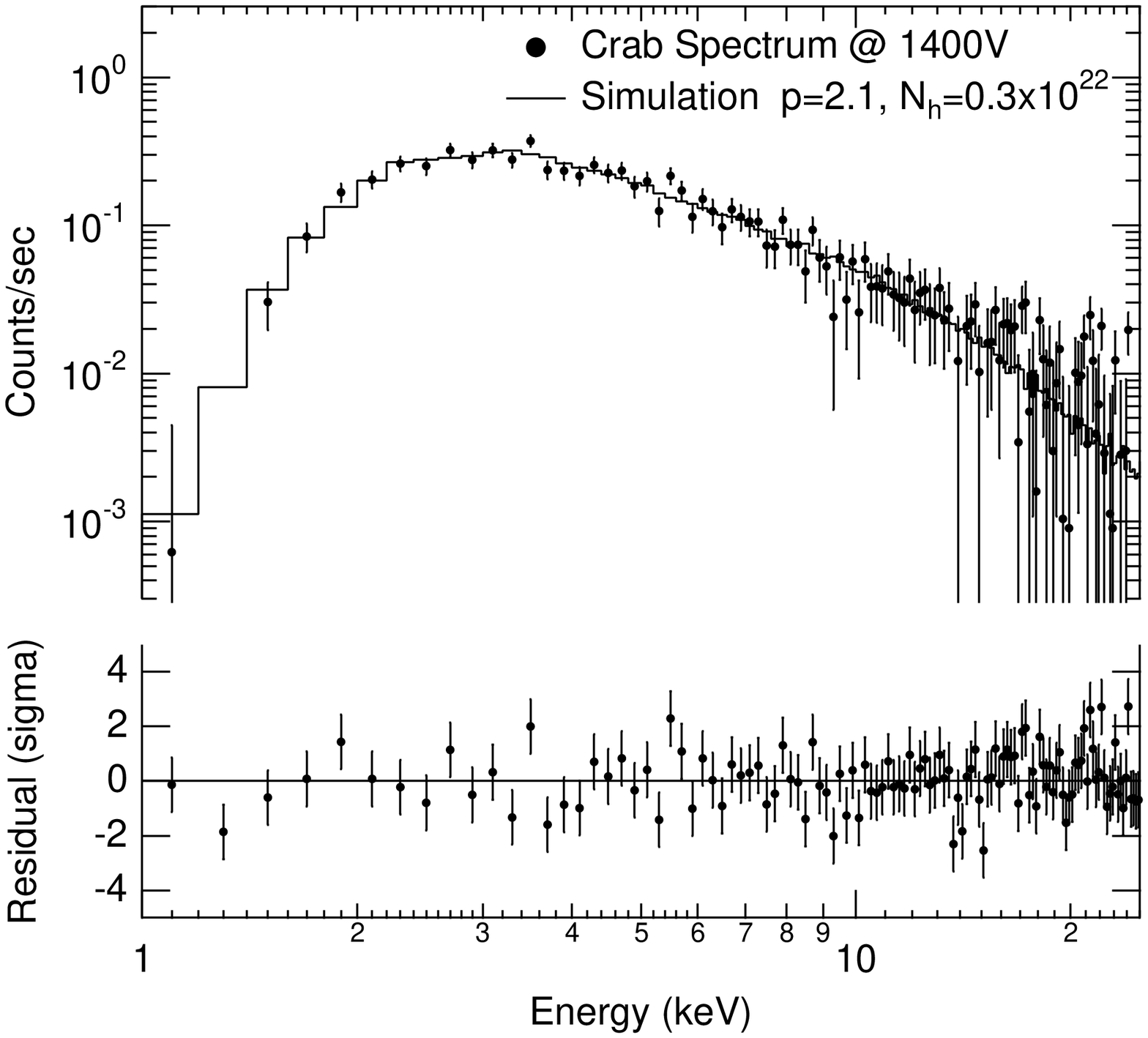}
         \caption{Comparison of the observed Crab Nebula spectrum with
         the simulated spectrum assuming a power-law with index 2.1 and 
         $3\times 10^{21}~\mbox{cm}^{-2}$ of absorption column density. The
         observation is performed at a bias voltage of 1400 V.}
         \label{fig:crab_spect_1400}
      }
      \parbox[t]{0.5\textwidth}{
         \FigureFile(0.5\textwidth,){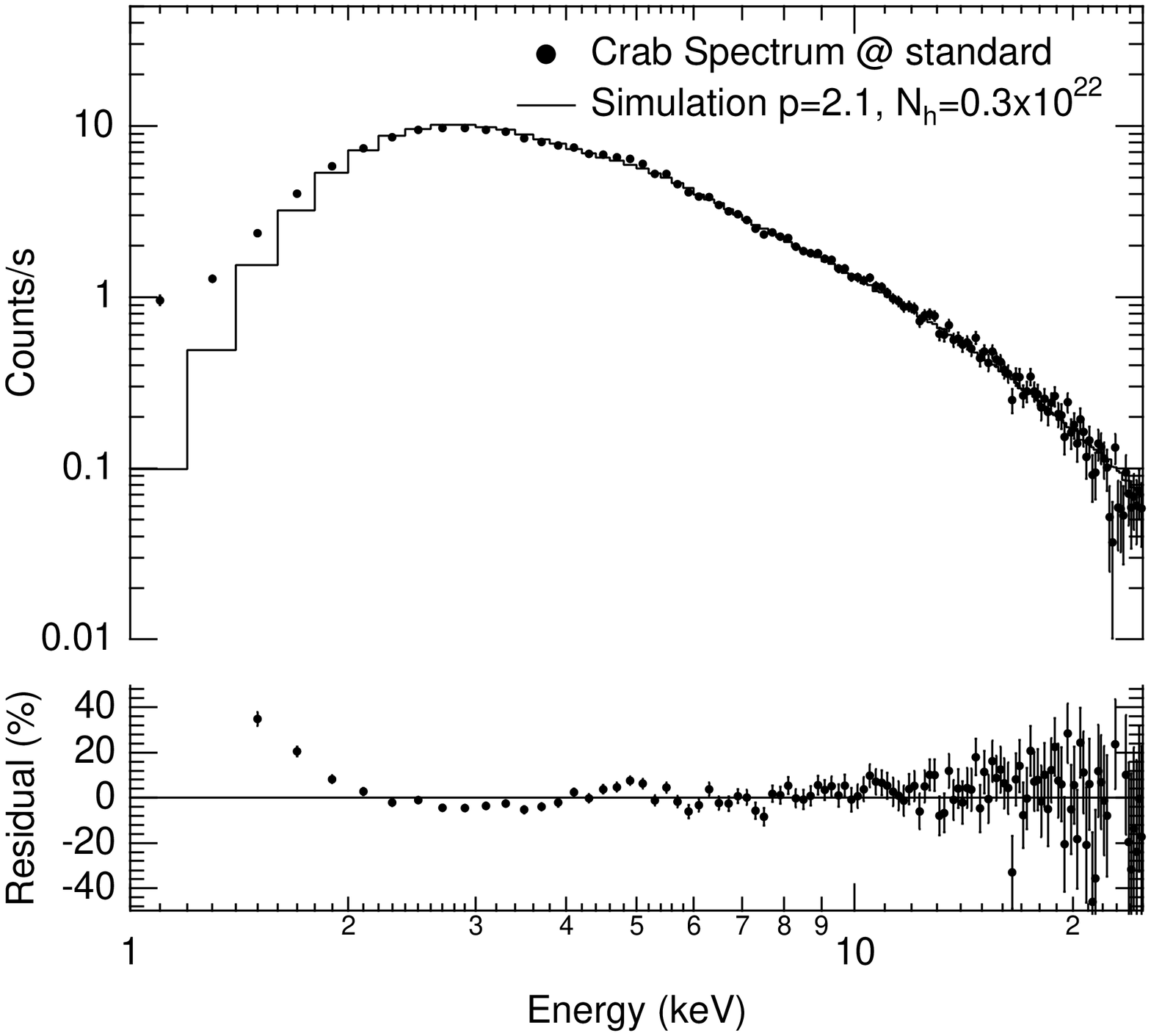}
         \caption{Same as figure~\protect\ref{fig:crab_spect_1400}, but
         for standard HV setting.}
         \label{fig:crab_spect_std}
      }
   }
\end{figure}

   To investigate the detector efficiency below 3~keV, we performed
observations of the Crab Nebula while operating at a low bias voltage
(1400~V).
   At this setting, the pulse height and energy relation shows better
linearity than at the standard setting, so we can check the detector
efficiency without being affected by uncertainty of their non-linear
relation at energies below 3~keV, where absorption by a Be film is
significant.
   The location of the Crab Nebula in instrument coordinates during the
observation period was $\theta_x = 21^{\circ}.6$ and $\theta_y =
12^{\circ}.7$.  Only the YB detector was
calibrated by this observation.
   In figure~\ref{fig:crab_spect_1400}, the observed Crab Nebula spectrum
is compared with a simulated spectrum of power-law index 2.1 and
absorption column density $3 \times 10^{21}~\mbox{cm}^{-2}$, which is a
canonical Crab spectrum.
   The absolute gain scale at 1400~V is determined by scaling down of the
effective bias voltage parameter for the standard HV setting so as to
reproduce the ground calibration data taken by $^{55}$Fe radioisotopes.
   The reduced $\chi^{2}$ of the comparison between the observed and
simulated spectra is 1.03 (d.o.f. = 118).
   From this result the accuracy of the detector efficiency at energies
below $\sim$ 4 keV is expected to be better than 10\%.

   Next, we present the result of Crab Nebula observations carried out
at the standard setting, that is 1668 V for XA and YB, and 1653 V for XB
and YA.
   The gain scale is determined by adjusting the gain parameters so as to
reproduce the canonical Crab Nebula spectrum for each wire.  The
measured energies of the calibration sources are checked to be accurate
within 2\%.
   In figure~\ref{fig:crab_spect_std}, the observed Crab Nebula spectrum
is compared with the simulated spectrum.
   Above 2 keV, the reduced $\chi^{2}$ between the two spectra is 1.79 
(d.o.f. = 114), so they are statistically inconsistent with each other.
   The most significant inconsistency is the spectrum around 5 keV,
where the observed spectrum is 5\% higher than the Monte Carlo
simulation.
   The cause of this inconsistency is under investigation.
   Below 2 keV, the observed count rates are higher than the expected
ones.
   This is probably due to the inaccuracy in the energy-pulse height
relation, and/or energy resolution at energies below 2 keV, where we
have no calibration data.

   Spectrum fitting for Crab Nebula observations was performed for
various incidence angles~\citep{wxm.SPIE.2003}.
   From this result, the systematic uncertainty of the power-law index,
absorption column density, and absolute flux are estimated to be
$\pm 0.1$, $\pm 0.2\times 10^{22}~$cm$^{-2}$ and $\pm$10\%, respectively.

   \subsection{WXM Alignment Calibration}

\begin{figure}
   \centerline{\FigureFile(0.6\textwidth,){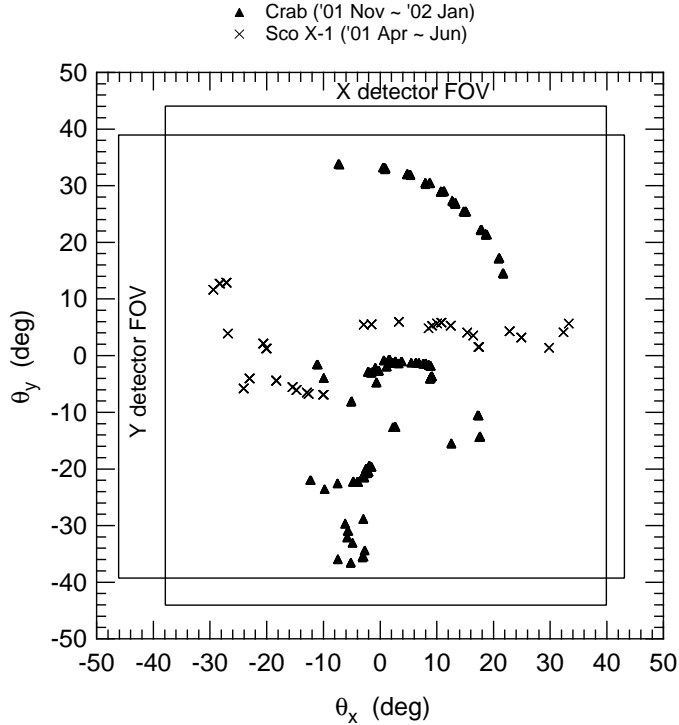}}
   \caption{Locations of Crab and Sco X-1 during the WXM alignment
 calibration.}
   \label{fig:crab-sco-location}
\end{figure}

   To calibrate the alignment between the WXM optical axis and the
spacecraft Z-axis, we used observations of the Crab Nebula and of Sco X-1.
   The observation periods of the Crab Nebula and Sco X-1 are given in
table~\ref{tbl:obs_period}.  The source locations in the WXM field
of view are plotted in figure~\ref{fig:crab-sco-location}.
   These observations were made in the quiet background region of the
HETE-2 orbit.
   The region of $\theta_x = -30^{\circ}$ to $33^{\circ}$ and $\theta_y
= -37^{\circ}$ to $34^{\circ}$ are calibrated by these observations.

   The data were divided into 20~s and 50~s
subsets for Sco X-1 and Crab Nebula data, respectively, and the localization
procedure was applied to each subset.
   The data type used in this analysis was RAW.  The charge ratio (PM)
correction derived in subsection~\ref{sec:flight_pos_cal} was applied as
a function of UT.
   The binning size of the position histograms was 0.2 mm.
   The background was obtained from the data taken in the quiet background
orbital region, during periods when Crab Nebula and Sco X-1 were out of the field 
of view.
   The source locations in the WXM coordinate system were calculated by
the cross-correlation method, as described in section~\ref{sec:riken_method}.
   In simulating the position histogram used for taking a
cross-correlation with the observed data, a power-law energy spectrum
of power index 2.1 was assumed for the Crab Nebula and a 6 keV thermal
bremsstrahlung spectrum was assumed for  Sco~X-1.

   The calculated directions in WXM coordinates were translated to
the spacecraft coordinates system using the rotation matrix, defined as
\begin{equation}
   V_{\mbox{\scriptsize SC}} = R_{x}(a) \cdot R_{y}(b) \cdot R_{z}(c)
                               \cdot V_{\mbox{\scriptsize WXM}},
   \label{eq:rotation}
\end{equation}
   where $V_{\mbox{\scriptsize WXM}}$ is the unit vector describing the
source location in WXM coordinates, $V_{\mbox{\scriptsize SC}}$ is the
unit vector describing the source location in spacecraft coordinates,
$R_{x}(a)$ is a rotation matrix with respect of the $x$-axis, and so on.
   The location in celestial coordinates was obtained by using the
spacecraft aspect data, expressed as the direction of the spacecraft
Z-axis in celestial coordinates and the roll angle about the $Z$ axis.
   The precision of the spacecraft aspect used in this analysis was
typically 1 arcmin.
   The Euler angles ($a$, $b$ and $c$) were determined by least-squares
minimization of the difference between the calculated locations and
the actual source locations.
   We defined $\chi^{2}$ as
\begin{equation}
   \chi^{2} = \sum \left\{ 
                   \frac{(\theta^{\mbox{\scriptsize cal}}_{x,i} 
                         -\theta^{\mbox{\scriptsize src}}_{x,i})^2}
                        {\sigma_{x,i}^{2}}
                 + \frac{(\theta^{\mbox{\scriptsize cal}}_{y,i} 
                         -\theta^{\mbox{\scriptsize src}}_{y,i})^2}
                        {\sigma_{y,i}^{2}}
                   \right\},
\end{equation}    
where $\theta^{\mbox{\scriptsize cal}}_{x,i}$ and
$\theta^{\mbox{\scriptsize cal}}_{y,i}$ are the $x$ and $y$ components of
the calculated location for the $i$-th sample, $\theta^{\mbox{\scriptsize
src}}_{x,i}$ and $\theta^{\mbox{\scriptsize src}}_{y,i}$ are the actual
source location derived from the spacecraft aspect data and assumed
Euler angles, and $\sigma_{x,i}^2$ and $\sigma_{y,i}^2$ are
statistical localization errors for their respective components.
   The obtained rotation angles are $a = 0^{\circ}.18$, $b =
0^{\circ}.34$ and $c = 0^{\circ}.07$.
   The r.m.s. of the localization error of Sco~X-1 is reduced from
20.0$'$ to 2.8$'$ by the alignment correction, and the r.m.s. for the
Crab Nebula is 4.3$'$ after the correction.

\begin{figure}
   \parbox{\textwidth}{
      \FigureFile(0.5\textwidth,){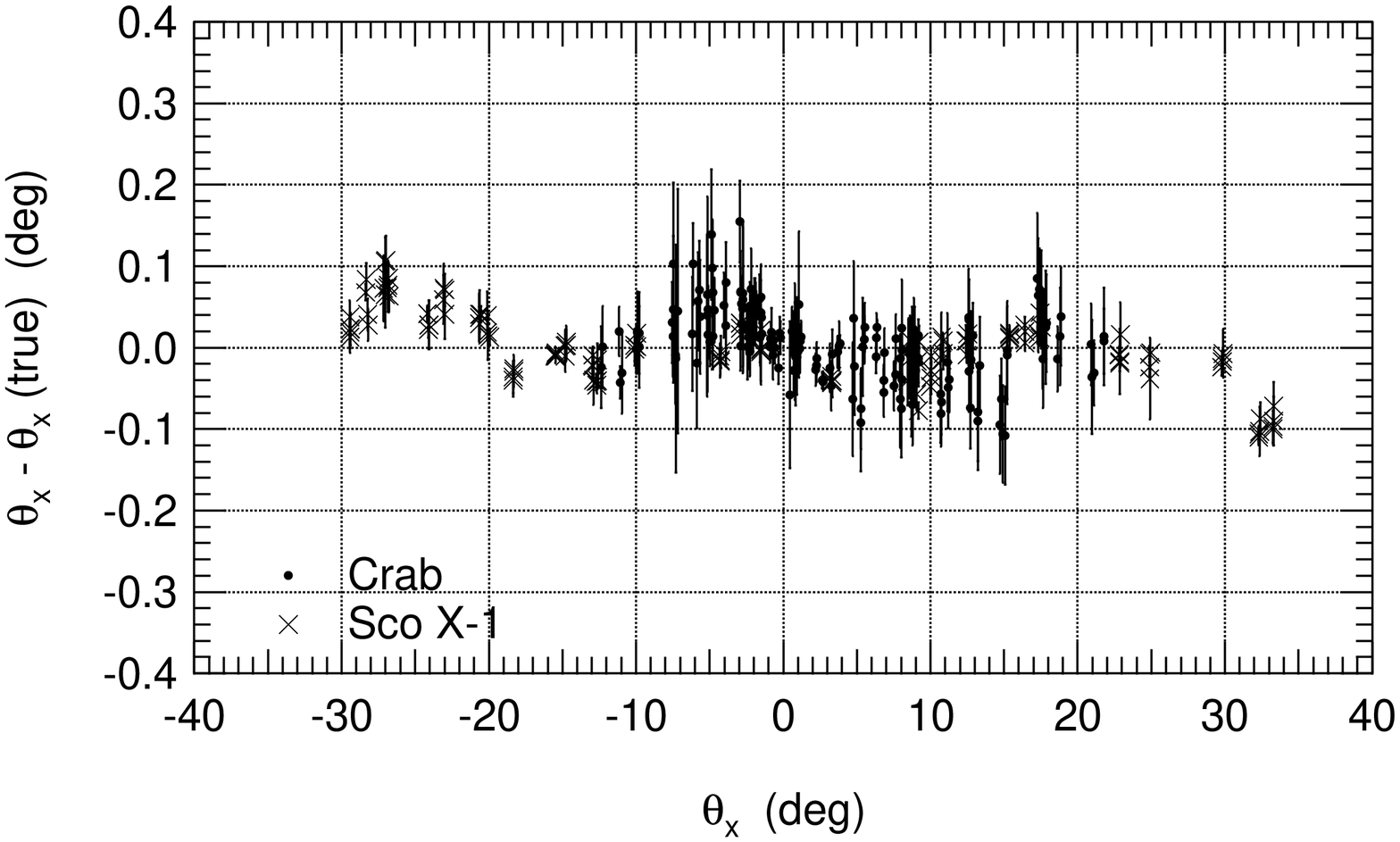}
      \FigureFile(0.5\textwidth,){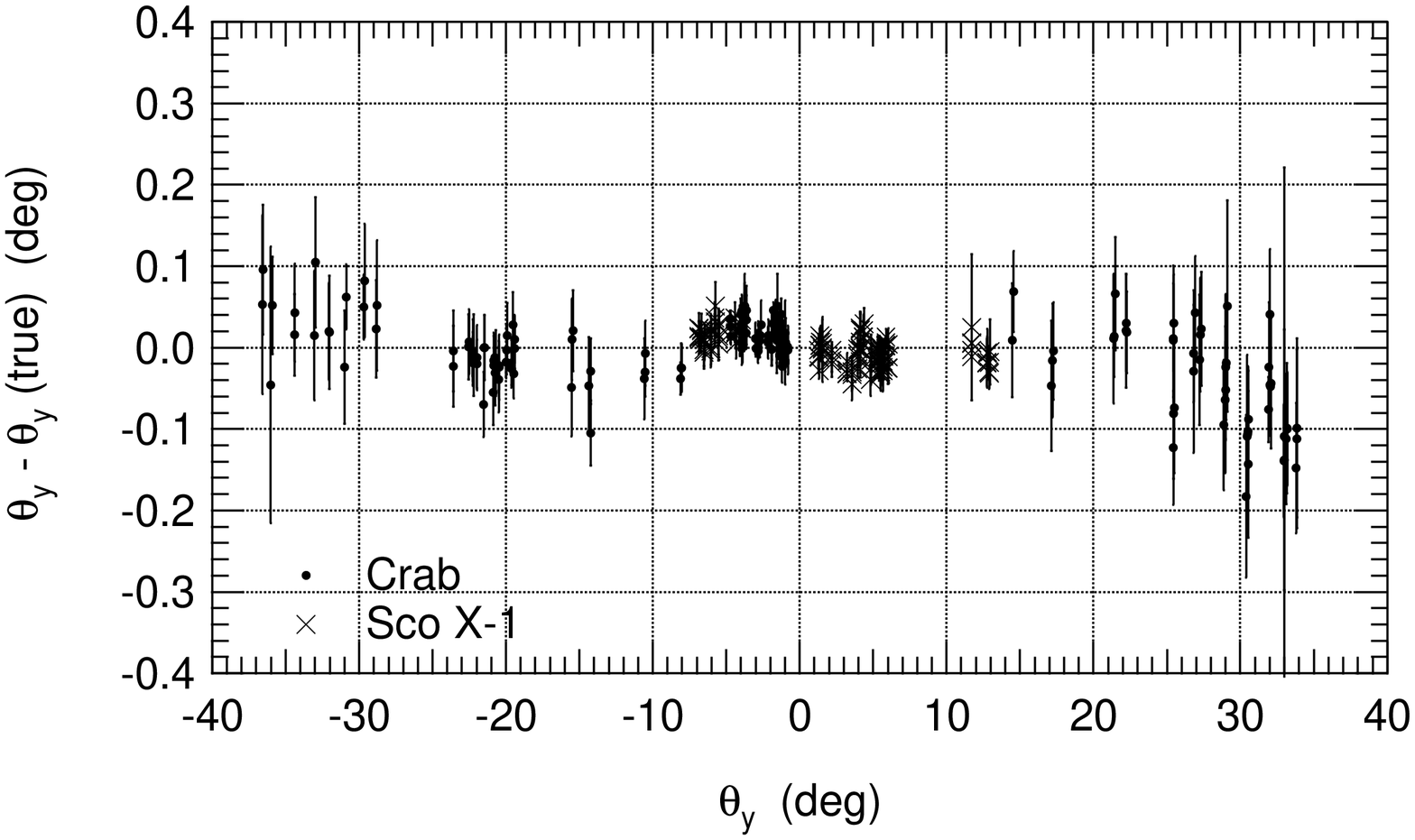}
  }
  \caption{Localization errors plotted against the source direction
           $\theta_{x}$ and $\theta_{y}$.}
  \label{fig:error_vs_thxy}
\end{figure}

   In figure~\ref{fig:error_vs_thxy} the localization error of $\theta_x$
and $\theta_y$ are plotted against $\theta_x$ and $\theta_y$, respectively.
   The errors of the directions are less than 0$^{\circ}$.1 at
any calibration region and no significant directional dependence is seen.
   The overall systematic error is estimated by requiring that the reduced
$\chi^{2}$ be unity by adding an additional error term in quadrature
to the statistical error.
   The resulting systematic error is $1'.83 \pm 0'.14$, to be added in
quadrature to the statistical error in each of the
$\theta_{x}$ and $\theta_{y}$ directions.
\begin{table}
\begin{center} 
   \caption{Observational periods of Crab Nebula and Sco~X-1 for the
            purpose of alignment calibration.}
   \label{tbl:obs_period}
   \begin{tabular}{ccc} \hline \hline
   Source      & Number of samples & Period \\ \hline
   Crab        & 437           & 2001 Nov. 08 -- 2002 Jan. 19 \\ 
   Sco X-1     & 208           & 2001 Apr. 23 -- 2001 Jun. 30 \\ \hline
   \end{tabular}
\end{center}
\end{table}

   We have investigated the accuracy of the estimated localization error
radius using the XRB and SGR events, whose coordinate are known to better
precision than the WXM localization precision.
   The error is uniform in the location in the FOV, and no significant
systematic deviation from the true location is seen for any direction.
   The error distribution of XRBs follows the expected
distribution, confirming that the error radius is correctly
estimated.
   More details are described in \citet{wxm.SPIE.2003}.

\subsection{Performance of In-Orbit GRB Localization and Operational Efficiency}

\begin{figure}
   \centerline{\FigureFile(0.8\textwidth,){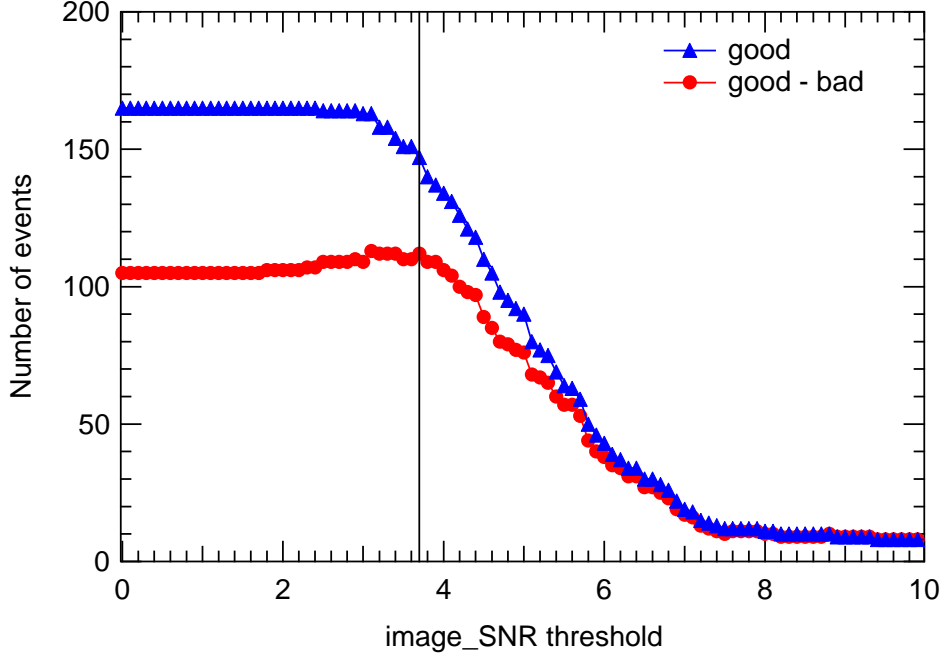}}
   \caption{Number of events localized to known X-ray sources with
            an error less than 0$^{\circ}$.5 (good events) are shown with
            triangles as a function of the threshold of imaging SNR,
            where imaging SNR is defined by a square root of sum in
            quadrature of image SNRs for X and Y directions.
            The differences between the number of good events and bad
            events (error larger than or equal to 0$^{\circ}$.5) are
            shown with solid circles.
            }
   \label{fig:good_bad}
\end{figure}

   The accuracy of the flight localization can be inferred from the 
total cross-correlation score, i.e. imaging SNR defined by 
$\sqrt{\mbox{IMAGE\_SN\_X}^{2} + \mbox{IMAGE\_SN\_Y}^2}$,
where IMAGE\_SN\_X,Y represent the cross-correlation score
corresponding to the X and Y directions, respectively.
   In order to determine a threshold criterion for automatically sending
out WXM localizations to the GRB Coordinates Network (GCN), we examined
the localization accuracy as a function of the imaging SNR.
   Figure~\ref{fig:good_bad} shows the number of correct localizations of
XRBs (good localization with location error less than 0$^{\circ}$.5) for which the
imaging SNR exceeds a given threshold value, and also shows the
the number of good localizations minus the number of bad localizations 
(localization error larger than 0$^{\circ}$.5).
  The criteria for sending to the GCN should be determined so that
the false alert rate is minimized while the number of correct
localizations is maximized.
  As a compromise between the two requirements, we chose a threshold
of the imaging SNR such that the difference between the number of good 
and bad localization is maximal.
  From figure~\ref{fig:good_bad}, the threshold is 3.7.

\begin{figure}
   \centerline{\FigureFile(0.8\textwidth,){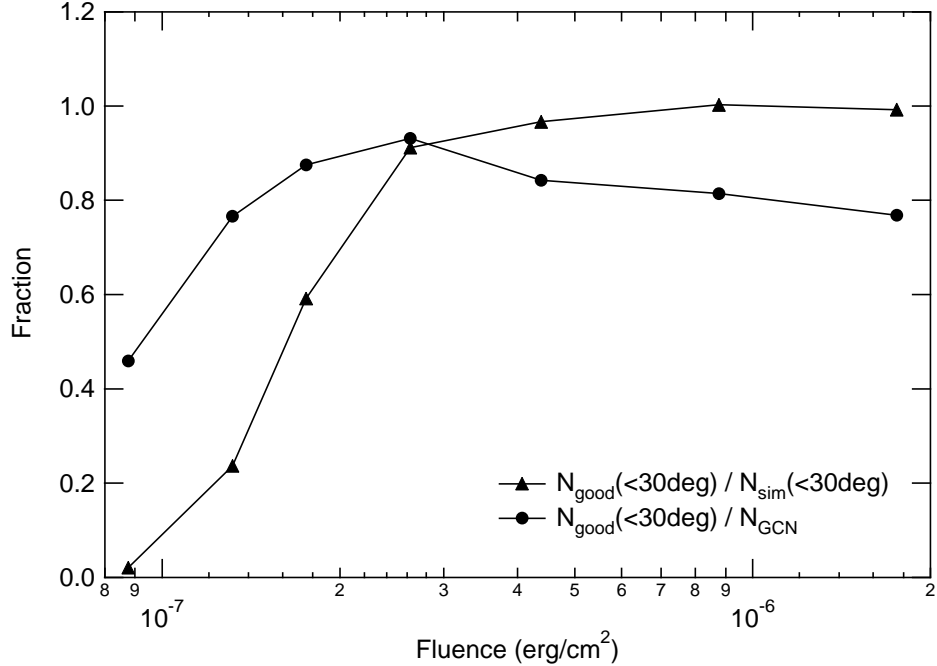}}
   \caption{Result from a flight localization simulation.
            The solid triangles show the fraction of correct
            localizations for events with incidence angles
            in the range $\theta_{x,y} = -30^{\circ}$ to $30^{\circ}$,
            and for which GCN criterion 2, i.e. image SNR larger than 3.7, is
            satisfied.
            The solid circles show the fraction of correct
            localizations for events with the calculated
            incidence angle in the range $\theta_{x,y}
            = -30^{\circ}$ to $30^{\circ}$, and for which GCN criterion
            2 is satisfied.
            }
   \label{fig:flight_loc_sim}
\end{figure}

  In order to estimate the sensitivity of the flight localization 
algorithm to the total fluence of a GRB, we carried out a
Monte Carlo simulation of GRB events, and applied the flight localization procedure
to the simulated data.
   In the simulation, we assumed
an energy spectrum of $E^{-1.5}$,  a uniform directional distribution in
the WXM FOV, a 10~s duration with a constant flux, and a background rate of
530 c~s$^{-1}$.
   The correct localization rate for the GRB locations in the range 
$\theta_{x,y} = -30^{\circ}$ to  $ 30^{\circ}$ (solid
triangles) and the fraction of correct localizations to be sent to the
GCN (solid circles) are shown in figure~\ref{fig:flight_loc_sim} as a
function of the total fluence.
   From this result, one can see that the flight localization
sensitivity is $> 10^{-7}$ erg~cm$^{-2}$, and that the false alert
rate is 50\% for GRB with $10^{-7}$ erg~cm$^{-2}$, reaches
maximum at $3 \times 10^{-7}$ erg~cm$^{-2}$, and decreases above that
fluence.
   The reason for the increase in the false alert rate above $3 \times 10^{-7}$
erg~cm$^{-2}$ is that GRB with locations at the edge of the FOV
($|\theta_{x,y}| > 30^{\circ}$) are incorrectly localized to the region
$|\theta_{x,y}| < 30^{\circ}$ and produce imaging SNR larger than 
3.7.

   Using $\sim$80 XRBs, we checked the statistics of the propagation times
of burst locations determined by the flight software~\citep{rome.2003.yuji}.
   The result shows that 30\% of the events are propagated to the ground
in 60~s, 50\% in 150~s, and 90\% in 590~s.
   The size of the localization error depends on the fluence and the
position of GRB in the FOV.  The typical total error ranges from 4
arcmin to 40 arcmin in diameter.

\begin{figure}
   \centerline{\FigureFile(1.0\textwidth,){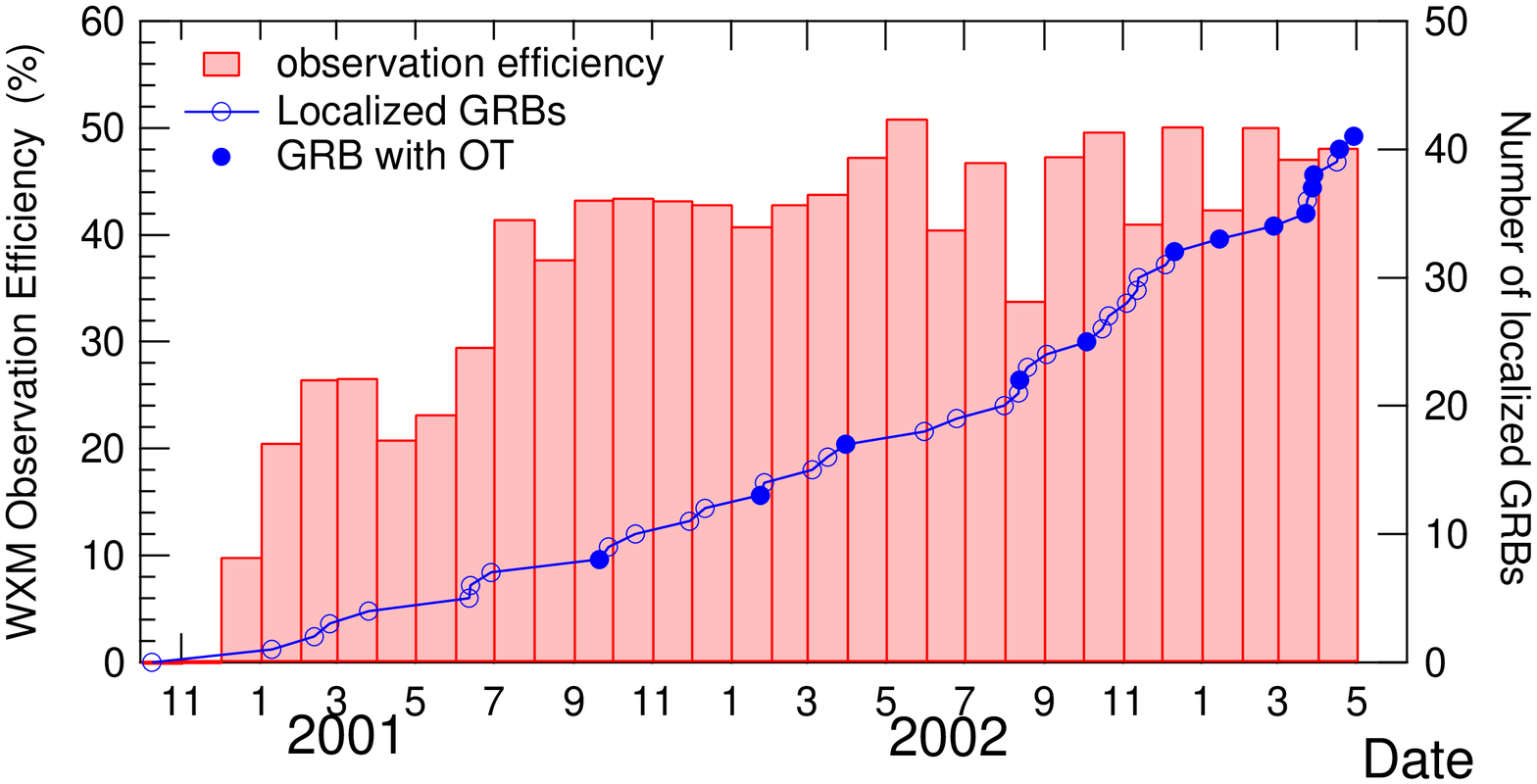}}
   \caption{Monthly operational efficiency and cumulative number of 
            localized
            GRBs. The efficiency is defined by a fraction of the total 
            HV on time in a month to the total time of the month.
            The GRBs for which optical transients are discovered are
            represented by solid circles.
            }
            \label{fig:ops_efficiency}
\end{figure}

   WXM operations are performed on the night side of the HETE-2 orbit.
   Operations are extended by several minutes beyond the terminators
during moderate solar activity, in order to enhance the GRB detection
probability.
   When the solar activity is high and X-class flares are likely, operations
are reduced to the dusk-to-dawn portion of the orbit, so as to avoid any
risk of direct irradiation by a strong solar flare in the event that a
loss of the spacecraft attitude control should result in WXM pointing toward
the Sun.
   The monthly operational efficiency is shown in
figure~\ref{fig:ops_efficiency},
together with the cumulative number of localized GRBs.
   On-time has been quite stable at a level of 40\%--50\% since
2001 September.  A total of 38 GRBs have been localized by the WXM as of 
2003 May 1.
   As seen in figure~\ref{fig:ops_efficiency}, the localization rate
increased around 2002 August, mainly due to
improvements in operational practice (principally head-nodding of the spacecraft to avoid
the moon and bright X-ray sources).
   The current estimated GRB localization rate is 28 per year, using
observations from 2002 August until 2003 April.

   In table~\ref{tbl:WXM-GRBs}, a summary of the localized GRBs, the delay
time for reporting their locations, dimensions of the error region quoted
in the first and last GCN notice/circular, and the approximate fluence of
the GRB are presented.
   Six real-time localizations were obtained  and propagated in times
ranging from 22~s to 9~min from the onset of the bursts.
   The typical delay time of the ground localization reports are 1 to
3 hr.
   The minimum fluence of the localized GRBs is $1\times
10^{-7}$~erg~cm$^{-2}$ in 2--25 keV.

\begin{table}   
   \begin{center}
   \caption{Summary of the localized GRBs.}
   \label{tbl:WXM-GRBs}
   \begin{tabular}{lp{4em}p{5em}p{4em}p{5em}p{9em}p{10em}} \hline \hline
                   & Reported time         & On board delay
                                           & Error size
                                           & Refined error size
                                           & Fluence (2--25 keV) (erg~cm$^{-2}$) 
                                           & Comment
                                           \\ \hline
       GRB 010213   & 36h14m &
                            & $60'$ dia.            
                            &
                            & $7\times 10^{-7}$ 
                            & ground trigger \\
       GRB 010326B  &  4h45m &
                            & $36'$ dia.          
                            & 
                            & $1\times 10^{-7}$ 
                            & \\
       GRB 010612   & 69h19m &
                            & $72'$ dia.   
                            &
                            & $5\times 10^{-7}$ 
                            & Sco X-1 in the FOV\\
       GRB 010613   & 56h55m &
                            & $72'$ dia.
                            &
                            & $6\times 10^{-6}$
                            & Sco X-1 in the FOV\\
       GRB 010629   &  8h57m &
                            & $30'$ dia.           
                            &
                            & $2\times 10^{-6}$ 
                            & Sco X-1 in the FOV\\
       GRB 010921   &  5h10m &
                            & $20'\times 10^{\circ}$
                            &
                            & $1\times 10^{-5}$  
                            & 1-D localization, OT \\
       GRB 010928   &  6h10m &
                            & $12'\times 10^{\circ}$
                            &
                            & $2\times 10^{-6}$  
                            & 1-D localization \\
       GRB 011019   & 12h07m &
                            & $70'$ dia.            
                            &
                            & $4\times 10^{-7}$  
                            & ground trigger \\
       GRB 011130   &  4h33m &
                            & $120'$ dia.           
                            & $15.2'$ dia.
                            & $6\times 10^{-7}$  
                            & \\
       GRB 011212   &  8h20m &
                            & $22'$ dia.            
                            &
                            & $5\times 10^{-7}$  
                            & ground trigger \\
       GRB 020124   &  1h26m &
                            & $27'\times 35'$       
                            & $24'$ dia. 
                            & $2\times 10^{-6}$  
                            & OT \\
       GRB 020127   &  1h46m &
                            & $24'$ dia.            
                            & $16'$ dia. 
                            & $6\times 10^{-7}$  
                            & \\
       GRB 020305   &  9h56m &
                            & $50'$ dia.
                            &
                            & (*1)
                            & OT \\
       GRB 020317   & 52m36s &
                            & $60'$ dia.            
                            & $36'$ dia. 
                            & $1\times 10^{-7}$  
                            & \\
       GRB 020331   & 40m22s &
                            & $20'$ dia.
                            & $16'$ dia.
                            & $1\times 10^{-6}$  
                            & OT \\
       GRB 020531   &  1h28m &
                            & $120'$ dia.           
                            & $43'\times 64'$
                            & $1\times 10^{-7}$  
                            & \\
       GRB 020625   &  2h54m &
                            & $18'\times 32'$       
                            & $18'\times 32'$
                            & $2\times 10^{-7}$  
                            & \\
       GRB 020801   &  1h56m &
                            & $11'\times 13^{\circ}$
                            & $9'\times 26'$ 
                            & $6\times 10^{-6}$ 
                            & \\
       GRB 020812   &  9m04s & 32 s
                            & $60'$ dia.            
                            & $10'\times 25'$
                            & $5\times 10^{-7}$ 
                            & real time alert \\
       GRB 020813   &  4m13s & 43 s
                            & $28'$ dia.            
                            & $8'$ dia. 
                            & $1\times 10^{-5}$  
                            & real time alert, OT \\
       GRB 020819   &  1h38m & 
                            & $14'$ dia.            
                            &                         
                            & $2\times 10^{-6}$ 
                            & \\
       GRB 020903   &  3h51m &
                            & $33'$ dia.            
                            &                         
                            & $1\times 10^{-7}$  
                            & \\
       GRB 021004   &    48s & 20 s
                            & $60'$ dia.            
                            & $20'$ dia.
                            & $5\times 10^{-7}$  
                            & real time alert, OT \\
       GRB 021016   &  1h44m &
                            & $19'\times 6^{\circ}$ 
                            & $15'\times 11^{\circ}$
                            & $3\times 10^{-6}$  
                            & 1-D localization \\
       GRB 021021   & 16h15m &
                            & $40'$ dia.            
                            &                        
                            & $1\times 10^{-7}$  
                            & ground trigger\\  
       GRB 021104   &  2h46m &
                            & $25'\times 42'$       
                            &                         
                            & $4\times 10^{-7}$  
                            & \\  
       GRB 021112   &  1h21m &
                            & $54'$ dia.            
                            & $40'$ dia.
                            & $2\times 10^{-7}$  
                            & \\
       GRB 021113   &  2h01m &
                            & $14'\times 29'$       
                            & $9'\times 26'$
                            & $1\times 10^{-6}$  
                            & \\
       GRB 021211   &    22s & 5.5 s
                            & $28'$ dia.            
                            & $10'$ dia. 
                            & $1\times 10^{-6}$  
                            & real time alert, OT \\  
       GRB 030115   & 1h11m  &
                            & $20'$ dia.
                            & 
                            & $8\times 10^{-7}$ 
                            & OT \\
       GRB 030226   & 1h49m  &
                            & $30'$ dia.
                            & 
                            & $2\times 10^{-6}$  
                            & ground trigger, OT \\
       GRB 030323   &  4h58m &
                            & $36'$  dia.
                            & 
                            & $4\times 10^{-7}$  
                            & OT \\
       GRB 030324   &  24.2s & 19 s
                            & $28'$ dia.
                            & $14'$ dia.
                            & $5\times 10^{-7}$  
                            & real time alert\\
       GRB 030328   &  2h01m &
                            & $9'$ dia.
                            & 
                            & $4\times 10^{-5}$  
                            & OT \\
       GRB 030329   &  (*2)  &
                            & (*2)
                            & $12'\times 2.25^{\circ}$
                            & $5\times 10^{-5}$  
                            & 1-D localization, OT \\
       GRB 030416   & 15h44m &
                            & $14'$ dia.
                            &
                            & $1\times 10^{-6}$ 
                            & ground trigger\\
       GRB 030418   &  3m35s & 31 s
                            & $28'$ dia.
                            & $18'$ dia.
                            & $2\times 10^{-6}$ 
                            & real time alert, OT \\
       GRB 030429   &  1h52m &
                            & $10'\times 20'$
                            & 
                            & $8 \times 10^{-7}$ 
                            & OT \\
                            \hline 
   \multicolumn{7}{c}{\parbox{\textwidth}{\footnotesize
          Note. Given are delay time for reporting
the WXM location, delay time of the on-board location relative to the
GRB onset time, dimension of the error region quoted in the first and
the last GCN notice/circular, and approximate fluence in 2--25 keV.
(*1) The fluence cannot be estimated due to incomplete observation. (*2) The
WXM location was not reported at the first GCN notice, instead the SXC
location was reported.}\hss}
   \end{tabular}
   \end{center}
\end{table}

\section{Summary}
We have calibrated the alignment between the WXM and the HETE-2 spacecraft
aspect system.
We estimate the systematic uncertainty of WXM locations of GRB to 
be 2$'$ for each X and Y direction at the 68\% level.  The  corresponding two-dimensional
systematic error is 4$'$ (90\% confidence level).
We calibrated the gain scale of the proportional counters using
observations of the Crab Nebula and using the signal from the calibration
radio-isotopes.
At present, the systematic uncertainties in fitting a power-law
spectrum model to the Crab spectrum are: $\pm 0.1$, $\pm 0.2\times 10^{22}$
cm$^{-2}$ and 10\% for power-law index, absorption column density, and
absolute flux, respectively.
Improvements in the spectral calibration are in progress.
The in-orbit flight localization performance of the WXM was examined based
on XRB observations and Monte Carlo simulations.
The sensitivity of the flight localization is estimated to be 
$> 10^{-7}$ erg~cm$^{-2}$, and the estimated false alert rate of real-time
localizations is 20\% for $1.5\times 10^{-7}$--$2 \times
10^{-6}$ erg~cm$^{-2}$.
Based on the recent GRB localizations, the rate of GRB localization
is 28 per year, and the real-time localization rate is approximately one per month.
%


\end{document}